\documentclass[letterpaper,prx,aps,floatfix,notilepage,nofootinbib,superscriptaddress,reprint,longbibliography]{revtex4-2}

\usepackage[english]{babel}
\usepackage{amsthm}
\usepackage{amssymb}
\usepackage{mathtools}

\usepackage{etoolbox}

\usepackage{graphicx}
\usepackage{dcolumn}
\usepackage{bm}
\usepackage{appendix}

\usepackage{xcolor}
\definecolor{teal}{HTML}{008080}

\usepackage[
    colorlinks=true,
	citecolor=teal,
	linkcolor=teal,
	urlcolor=teal]{hyperref}

\usepackage{subcaption}
\usepackage{xcolor}

\newcommand{\bts}[1]{{\bf{#1}}}
\newcommand{\ket}[1]{|#1\rangle}
\newcommand{\bra}[1]{\langle#1|}
\newcommand{\braket}[3]{\langle#1|#2|#3\rangle}
\newcommand{\overlap}[2]{\langle#1|#2\rangle}
\newcommand{\ntrot}{n_{\mathrm{T}}}
\newcommand{\nqubits}{n_{\mathrm{spin}}}
\newcommand{\meanwork}{\mathrm{E}[\mathcal{W}]}
\newcommand{\meanheat}{\mathrm{E}[\mathcal{Q}]}
\newcommand{\meaninten}{\mathrm{E}[\Delta \mathcal{U}]}
\newcommand{\varwork}{\mathrm{Var}[\mathcal{W}]}
\newcommand{\supp}[1]{Appendix #1}
\newcommand{\snr}{\mathrm{SNR}[\mathcal{W}]}
\newcommand{\entropy}{\Sigma}
\newcommand{\pauli}[1]{{\mathsf{#1}}}
\newcommand{\device}[1]{${\mathsf{ibm\_#1}}$}
\newcommand{\moment}[1]{m_{\mathcal{W},#1}}

\newcommand{\figuretitle}[1]{{\bf{#1}}}
\newcommand{\edge}[2]{(#1,#2)}
\newcommand{\library}[1]{$\mathsf{#1}$}
\newcommand{\gibbs}[2]{\pi_{#1}^{#2}}
\newcommand{\trace}[1]{\mbox{Tr}\left[#1\right]}
\newcommand{\ttmat}[4]{\left( \begin{array}{cc} #1 & #2 \\ #3 & #4 \end{array} \right)}

\begin{document}

\title{Non-equilibrium thermodynamics of precision \\ through a quantum-centric computation}

\author{Mario Motta}
\email{mario.motta@ibm.com}
\affiliation{IBM Quantum, IBM T. J. Watson Research Center, Yorktown Heights, NY 10598, USA}

\author{Antonio Mezzacapo}
\affiliation{IBM Quantum, IBM T. J. Watson Research Center, Yorktown Heights, NY 10598, USA}

\author{Giacomo Guarnieri}
\email{giacomo.guarnieri@unipv.com}
\affiliation{Department of Physics and INFN - Sezione di Pavia, University of Pavia, Via Bassi 6, 27100, Pavia, Italy}

\date{\today}

\begin{abstract}
Thermodynamic uncertainty relations (TURs) are a set of inequalities expressing a fundamental trade-off between precision and dissipation in non-equilibrium classical and quantum thermodynamic processes. TURs show that achieving low fluctuations in a thermodynamic quantity (e.g., heat or work) requires a minimum entropy production, with profound implications for the efficiency of biological and artificial thermodynamic processes.
The accurate evaluation of TURs is a necessary requirement to quantify the fluctuation and dissipation entailed by a thermodynamic process, and ultimately to optimize the performance of quantum devices, whose operation is fundamentally a quantum thermodynamic process.
Here, we simulate TURs in a transverse-field Ising model subjected to a time-dependent driving protocol, using quantum and classical computers in concert. 
Varying the duration and strength of the drive as well as the system size, we verify the validity of TURs, identify quantum signatures in the work statistics within the linear response regime, and observe TUR saturation in the high-temperature limit.
\end{abstract}

\begin{figure*}[t!]
\includegraphics[width=0.92\textwidth]{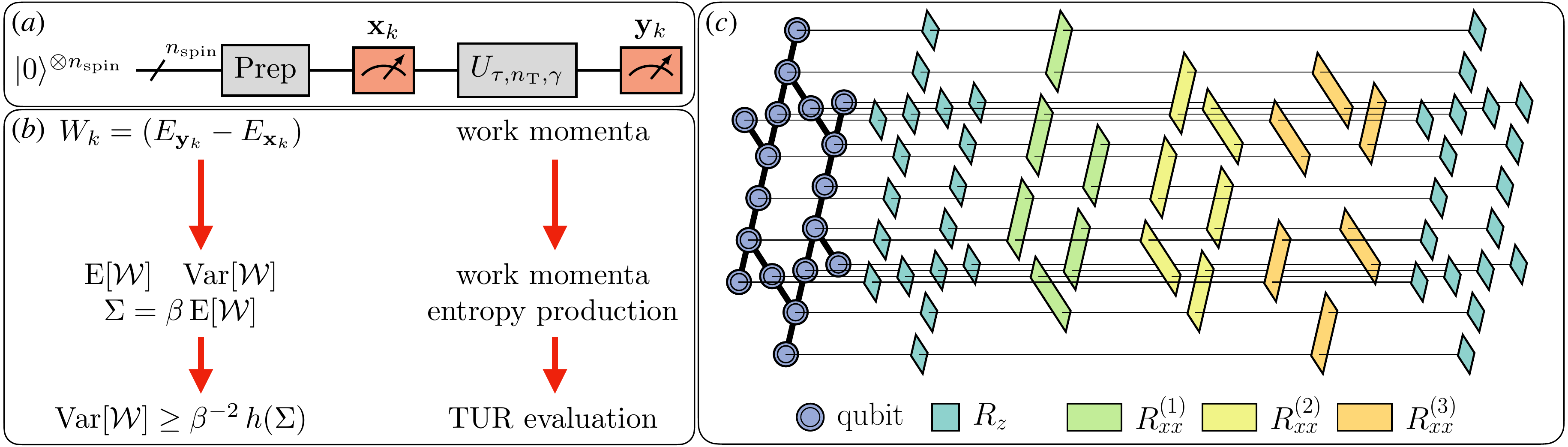}
\caption{\figuretitle{Drive protocol and quantum circuits.} 
(a) The protocol includes a state preparation operation (gray box, ``Prep'') and a quantum circuit simulating the action of a drive protocol (gray box, $U_{\tau,\ntrot,\gamma}$) surrounded by mid- and final-circuit measurements in the Pauli-$\pauli{Z}$ basis (red meters), yielding computational basis states $({\bf{y}}_k,{\bf{x}}_k)$.
(b) Computational basis states yield quantum work samples that we use to compute the mean, variance, and entropy production of the quantum work distribution, and verify experimentally a thermodynamic uncertainty relation (TUR).
(c) Quantum circuit implementing a step of time evolution in the drive protocol, comprising two layers of $\pauli{Z}$ rotations (teal squares) surrounding three layers of $\pauli{XX}$ rotations (green, yellow, and orange rectangles).
}
\label{fig:setup}
\end{figure*}

\maketitle 

Quantum devices, once the subject of theoretical speculation, are becoming increasingly available. Progress in their manufacturing and operation is poised to reshape diverse fields of technology, including computing, sensing, communication, and applied thermodynamics.
Their transformative potential is fundamentally dependent on the ability to harness genuinely quantum features (such as superposition, coherence, and entanglement) to outperform their classical counterparts in the execution of specific tasks. 

However, the operation of any quantum or classical device is a physical process, and therefore its capabilities are ultimately dictated and bound by the laws of physics~\cite{lloyd2000ultimate}.
Among these, thermodynamics plays a fundamental role, because operating a device requires to drive or maintain a system out of equilibrium through exchanges of work, heat, and/or other quantities with an environment. 
These processes are unavoidably accompanied by irreversible energy dissipation \cite{landauer1961irreversibility,bennett1973logical}, identified and quantified by the entropy production~\cite{LandiReview}.
Therefore, understanding and optimizing the energetic and entropic footprint of quantum devices is not just desirable; it is essential for their scalability and usefulness \cite{auffeves2022_QEI}.

However, the energy cost is only one piece of the puzzle: the precision of quantum technologies is also critical to achieve their full potential. Precision is quantified by fluctuations in measurable quantities that unavoidably occur in both the classical and quantum regimes~\cite{Feldmann,Plastina,Francica2017,Dann2019}. Characterizing the relationship between fluctuations and dissipation is necessary to quantify the thermodynamic cost of precise device operation, and is one of the central themes of stochastic thermodynamics \cite{kubo1986brownian,onsager1931reciprocal,kubo1966fluctuation,Gallavotti1995b,Jarz97a,Crooks1998,Tasa00a,Kurc00a,Jarz04a, Andrieux2009,Saito2008,Espo09a,Camp11a,Jarz11a,Hanggi2015}.

A major breakthrough in this field is the recent discovery of a collection of inequalities called thermodynamic uncertainty relations (TUR)~\cite{Barato2015,gingrich2016dissipation}. Originally discovered for steady states of Markov chains ~\cite{Barato2015,gingrich2016dissipation}, and later generalized to non-equilibrium classical ~\cite{Pietzonka2017a,Dechant2018,horowitz2017proof,proesmans2017discrete,BaratoNJP2018,Dechant2018,dechant2020fluctuation} and quantum~\cite{Holubec2018PRL,VanVu2020,Guarnieri2021PRL,Guarnieri2021PRE,falasco2020unifying,hasegawa2020quantum,hasegawa2021thermodynamic,VuSasa2022,Hasegawa2023,VuSaito2023,MacIeszczak2018,dechant2020fluctuation,Guar19c,SegalAgarwalla,GerryAgarwalla,LiuSegal2021PRE} processes, TURs have the general form~\cite{Horo20a}
\begin{equation}
\label{eq:tur_intro}
\varepsilon_{\mathcal{Q}} \equiv  
\frac{\mathrm{Var}[\mathcal{Q}]}{\mathrm{E}[\mathcal{Q}]^2}
\geq 
f\left(\entropy\right)
\;,
\end{equation}
where $\mathcal{Q}$ is any thermodynamic quantity (e.g. work or heat) and $f$ is a monotonically decreasing function~\cite{timpanaro2019,Hase19a} of the average entropy production $\entropy$.

TURs reveal a rich and striking trade-off between the precision $\varepsilon_{\mathcal{Q}}$ and the dissipation $\Sigma$ of a thermodynamic process, dictating that increasing the former also requires increasing the latter.
Quantum effects in particular play a non-trivial role in this balance, typically resulting in enhanced fluctuations of thermodynamic quantities~\cite{Guarnieri2021PRL,guarnieri2024generalized,Miller2020Landauer} and measurable signatures in the probability distribution of the entropy production~\cite{Miller2019,Scandi2019,abiuso2020geometric}.

TURs have immediate implications for the design and operation of quantum devices. As researchers strive to optimize quantum processes and devices for both performance and resource consumption~\cite{koch2022quantum,pekola2021colloquium, myers2022quantum,auffeves2022_QEI}, fundamental questions arise: what is the ultimate thermodynamic cost of performing a quantum process? Under what circumstances do quantum effects permit to outperform classical approaches when fluctuations are taken into account?

Numerical methods can support this emerging research by quantitatively evaluating $\varepsilon_{\mathcal{Q}}$ and $\entropy$ over particular thermodynamic processes, and ultimately optimizing such processes in the sense of Eq.~\eqref{eq:tur_intro} under constraints on fluctuations or dissipation. Computer simulations of non-equilibrium quantum thermodynamics are challenging and cutting-edge, as they require to capture the intricate interplay between quantum mechanics and thermodynamics in systems of many interacting particles. Quantum computers can provide a new path to simulate quantum thermodynamic processes~\cite{buffoni2020thermodynamics,solfanelli2021experimental,buffoni2022third,bassman2022computing,cimini2020experimental,cech2023thermodynamics,hahn2023quantum,Consiglio2024PRA}, especially when used in conjunction with classical computers~\cite{alexeev2024quantum} in the framework of hybrid quantum-classical algorithms.

Here, we simulate TURs for systems of 4 to 130 qubits using the Heron superconducting processor \device{torino} in concert with classical computing resources. We subject qubits to a driving protocol in which a pairwise interaction is modulated with an on-off cycle of tunable duration and strength. By measuring qubits before and after the pairwise interaction is engaged, we can access the quantum work distribution of the system, compute its first and second moments, and verify the validity of TURs as a function of the duration and strength of the driving protocol.
A technical advancement that enabled this simulation is a post-processing technique motivated by quantum selected configuration interaction~\cite{kanno2023qsci} and sample-based quantum diagonalization~\cite{robledomoreno2024}, where the thermodynamics of the system are projected in a subspace defined by computational basis states sampled from a quantum device and solved using linear algebra techniques on a classical device.

\section{Model}

The setup we consider is shown schematically in Fig.~\ref{fig:setup}. We consider the paradigmatic transverse-field Ising model (TFIM), described by the Hamiltonian
\begin{equation}
\label{eq:hamiltonian_main}
H_t = - \sum_{q \in V} Z_q + \frac{\lambda_t}{|E|} 
\sum_{\edge{p}{r} \in E} X_p X_r \;,
\end{equation}
where $(V,E)$ is a graph and $X,Y,Z$ denote the spin-$\frac{1}{2}$ Pauli matrices. The system starts in thermal equilibrium at an inverse temperature $\beta$ under the Hamiltonian $H_0$. The interaction between spins is adjusted by a protocol $\lambda_t = \gamma \, \sin(\pi t/\tau)$ acting for time $0 \leq t \leq \tau$, with a time-reversal symmetry $\lambda_{\tau-t} = \lambda_t$ leading to $H_0=H_\tau$. The driving protocol is simulated with a circuit of $\ntrot$ second-order Trotter-Suzuki~\cite{suzuki1990fractal} steps (see \supp{A}).

Measurements in the Pauli-$\pauli{Z}$ basis, performed at the start ($t=0$) and end $(t=\tau)$  of the protocol, yield an ensemble of two-point trajectories $(\bts{y}_k,\bts{x}_k)$, $k=1 \dots N$. Since the operators $H_0$ and $H_\tau$ are diagonal, over each trajectory we can compute a quantum work sample, $W_k = \langle \bts{y}_k | H_\tau | \bts{y}_k \rangle - \langle \bts{x}_k | H_0 | \bts{x}_k \rangle$, as the difference between the outcomes of the two projective measurements at the start and end of the protocol. Finally, the samples $W_k$ yield the mean $\meanwork$ and variance $\varwork$ of the quantum work distribution and, since the free energy difference is zero for this protocol, $- \beta \, \Delta F = \log \trace{e^{-\beta H_\tau}}/ \trace{e^{-\beta H_0}} = 0$, the mean work is proportional to the mean entropy production $\entropy = \beta \meanwork$.
These are the central quantities involved in the numerical evaluation of the TUR Eq.~\eqref{eq:tur_intro} in the form $\varwork \geq \beta^{-2} h(\entropy)$. Here and henceforth $h(\entropy) \equiv \Sigma^2 f(\Sigma)$ (see \supp{B1}).

In absence of noise, two-point trajectories follow the ideal distribution
\begin{equation}
\label{eq:ideal}
p_{\mathrm{id}}(\bts{y},\bts{x}) 
= 
| \langle \bts{y} | U^\tau_0 | \bts{x} \rangle|^2 \, \langle \bts{x} | \gibbs{0}{\beta} | \bts{x} \rangle,
\end{equation}
where $\gibbs{0}{\beta} = e^{-\beta H_0} / \mathrm{Tr}[e^{-\beta H_0}] $ is the equilibrium state at inverse temperature $\beta$ under the Hamiltonian $H_0$, and $U^\tau_0$ is the time evolution operator (approximated by the circuit $U_{\tau,\ntrot,\gamma}$ in Fig.~\ref{fig:setup}c). Noise, however, alters Eq.~\eqref{eq:ideal} and biases the statistics of work and other thermodynamic quantities. To mitigate these biases, we introduce a post-processing technique motivated by sample-based quantum diagonalization \cite{kanno2023qsci, robledomoreno2024}, where we project the Hamiltonian Eq.~\eqref{eq:hamiltonian_main} in the subspace $S$ spanned by the measured computational basis states (or a suitable extension~\cite{barison2024quantum} of it) and recompute the distribution using the ideal form Eq.~\eqref{eq:ideal} and the projected Hamiltonian $J_t = P_S H_t P_S$, where $P_S = \sum_{\bts{z} \in S} | \bts{z} \rangle \langle \bts{z} |$. We call this technique SQT (sample-based quantum thermodynamics) or Ext-SQT when employing $S$ or its extension (see \supp{B2}).

\section{Results}

\begin{figure}[t!]
\includegraphics[width=\columnwidth]{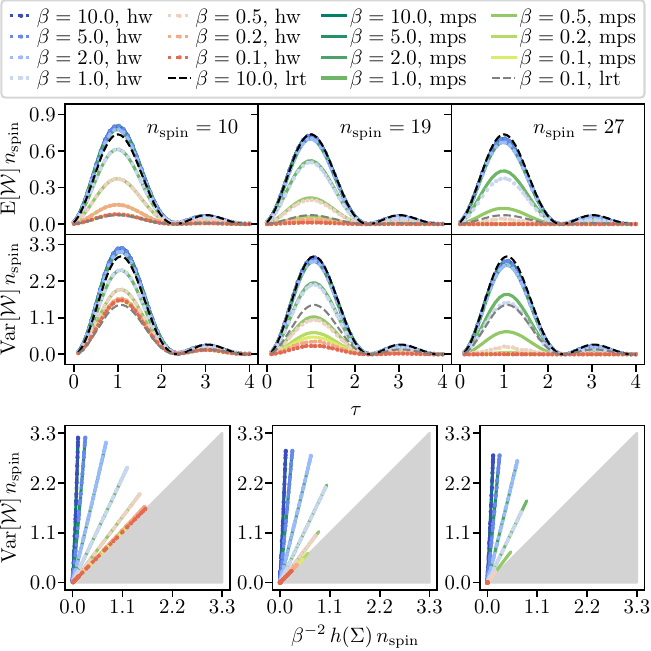}
\caption{\figuretitle{Thermodynamic uncertainty relations as a function of protocol duration.} First and second row: mean $\meanwork$ and variance $\varwork$ of the quantum work distribution as a function of the protocol duration $\tau$ computed on \device{torino} (dotted curves, ``hw'') and with matrix product states (solid curves, ``mps'') for $\nqubits = 10, 19, 27$ qubits and inverse temperatures from $\beta = 10.0$ to $\beta=0.1$ (blue to red for ``hw'' data and dark to light green for ``mps'' data, respectively). Third row: TUR from the data in the first and second row, with violations occurring in the gray region.}
\label{fig:scan}
\end{figure}

Fig.~\ref{fig:scan} shows the work statistics and thermodynamic uncertainty relations as a function of the duration $\tau$ of the protocol, with the interaction strength fixed at $\gamma=1$ and a circuit of $\ntrot=10$ time-evolution steps. The average and variance of the work are roughly inversely proportional to the size of the system, decrease monotonically with increasing temperature, and feature a dominant and a satellite peak at $\tau \simeq 1.0$ and $\tau \simeq 3.0$ respectively. All these features are taken into account by a linear response theory (LRT) analysis based on recent results ~\cite{guarnieri2024generalized}, that predicts the expression $\moment{k} = \gamma^2 f_k(\beta) g(4\tau) / |E|$ for the cumulants of the work distribution, where $k$ is the order of the cumulant ($k=1,2$ for mean and variance, respectively) and $f_k$ and $g$ are suitable functions of the inverse temperature and duration of the protocol (see  \supp{B3} for derivation and details).

As a consequence of this relation, the graph of the function $\tau \mapsto \big( \beta^{-2} h(\entropy)(\tau), \varwork(\tau) \big)$ is a straight line with a slope depending on the inverse temperature, visible in the lower panel of Fig.~\ref{fig:scan}.
As the temperature $\beta^{-1}$ increases, this line rotates from the vertical axis to the bisector; in the high-temperature regime, the average entropy production $\Sigma = \beta \meanwork$ becomes very small for any value of the duration $\tau$ of the protocol. Consequently, the TUR~Eq.~\eqref{eq:tur_intro} reduces to the classical trade-off $f(\Sigma) \simeq 2/\Sigma$, originally found in ~\cite{Barato2015}, which provides a tight bound saturated in the infinite-temperature limit. This result, showing that the system acquires an effective classical description when the thermal energy $\beta^{-1}$ exceeds any other energy scale in the system, is fully confirmed by our simulations. 
Remarkably, our data clearly show quantum signatures at low temperature, manifested in deviations from the fluctuation-dissipation relation $\varwork = 2\beta \meanwork$ predicted by classical Kubo linear response theory~\cite{kubo1966fluctuation}.

\begin{figure*}[t!]
\includegraphics[width=\textwidth]{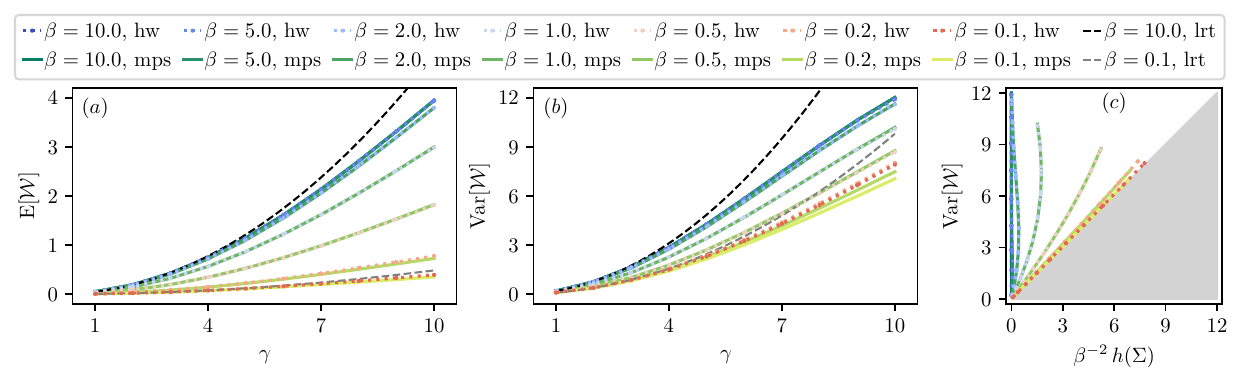}
\caption{\figuretitle{Thermodynamic uncertainty relation as a function of protocol strength.} Mean $\meanwork$ (panel $a$) and variance $\varwork$ (panel $b$) of the quantum work distribution as a function of the protocol strength $\gamma$ computed with \device{torino} (dotted curves, ``hw''), matrix product states (solid curves, ``mps''), and LRT (dashed curves, ``lrt'', black and gray for $\beta=10.0$ and $0.1$) for $\nqubits = 15$, $\ntrot = 8$ Trotter-Suzuki steps, and inverse temperatures from $\beta = 10.0$ to $\beta=0.1$ (from blue to red for ``hw'' data and from dark to light green for ``mps'' data, respectively). Panel $c$: TUR from the data in panels $a$ and $b$, with violations occurring in the gray region.
}
\label{fig:coup}
\end{figure*}

We also note that for small systems ($\nqubits = 10$) the data from tensor network (mps) and quantum computer (hw) samples post-processed with Ext-SQT are in agreement with each other. As the system size increases, differences between hw and mps data appear, reflecting the noise on quantum computing samples. Furthermore, hw and mps data differ from LRT (and exact numerical simulations, see \supp{D}) for higher temperatures and larger systems ($\beta < 1.0, \nqubits=27$). These differences reflect the breakdown of the sparsity approximation in the high-temperature limit, where $\gibbs{0}{\beta} \simeq I/2^{\nqubits}$. In this limit, the projection $P_S$ underlying sample-based post-processing biases the distribution Eq.~\eqref{eq:ideal} even for noiseless mps samples, to an extent that becomes more pronounced as system size increases. As a result Ext-SQT tends to underestimate the mean and variance of the work distribution, altering the length of the straight lines in the lower panel, though not their slopes.

Fig.~\ref{fig:coup} shows the work statistics and thermodynamic uncertainty relations as a function of the interaction strength $\gamma$, with $\nqubits=15$ and the duration of the protocol fixed at $\tau =1$ (close to the location of the dominant peak in Fig.~\ref{fig:scan}). Data from tensor network (mps) and quantum computer (hw) samples post-processed with Ext-SQT agree with each other (and with exact numerical simulations, see \supp{D}) for different values of $\gamma$. The mean and variance of the work distribution increase roughly quadratically with $\gamma$ as predicted by LRT (dashed lines), deviating from a parabolic dependence as the interaction strength increases past $\gamma \simeq 4$, particularly visible in the variance of the work (panel $b$). Consequently, the curves $\gamma \mapsto \big( \beta^{-2} h(\entropy)(\gamma), \varwork(\gamma) \big)$ in panel $c$ are no longer straight lines, except at very low and high temperatures. Note that, as in Fig.~\ref{fig:scan}, the TUR is saturated for a vanishing inverse temperature.

\begin{figure*}[t!]
\includegraphics[width=0.9\textwidth]{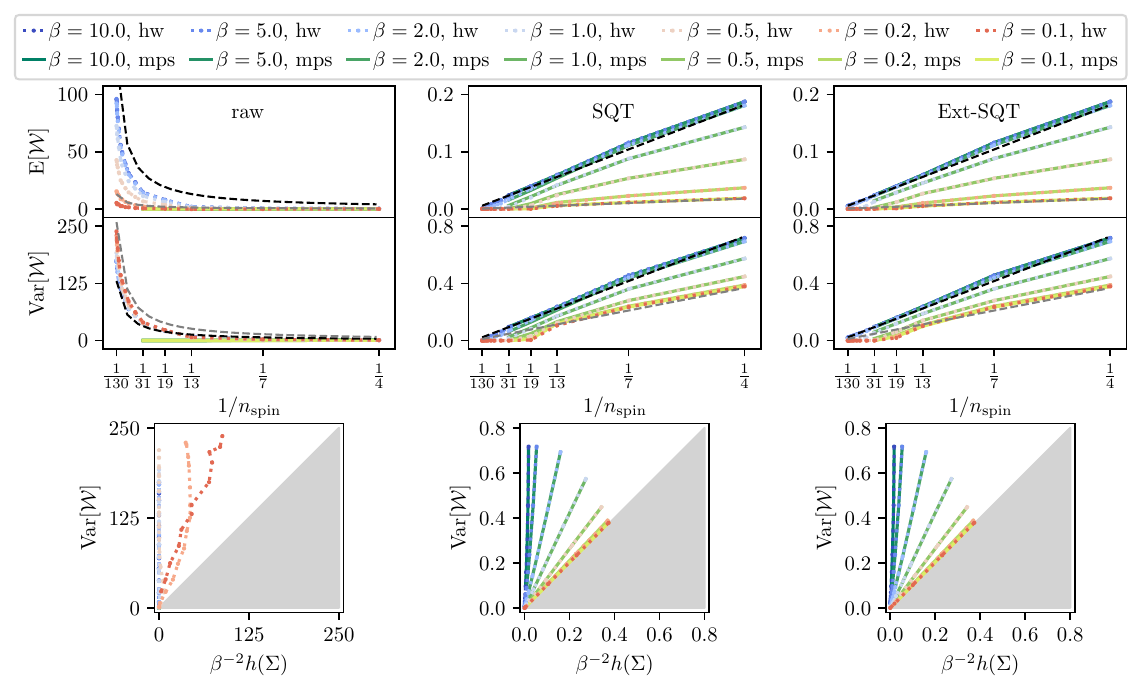}
\caption{\figuretitle{Thermodynamic uncertainty relation as a function system size.}
First and second row: mean $\meanwork$ and variance $\varwork$ of the quantum work distribution as a function of system size computed with \device{torino} (dotted curves, ``hw'') and matrix product states (solid curves, ``mps''), for $\nqubits = 4$ to $130$ qubits, up to $\ntrot=29$ Trotter-Suzuki steps, and inverse temperatures decreasing from $\beta = 10.0$ to $\beta=0.1$ (blue to red for ``hw'' data, dark to light green for ``mps'' data). Results are raw (left) and post-processed with ``SQT'' and ``Ext-SQT'' (center, right). Raw data are compared versus an extreme depolarizing noise model (black, gray dashed lines for $\beta=10.0$, $0.1$), and post-processed data versus LRT (black, gray dashed lines for $\beta=10.0$, $0.1$).
Third row: thermodynamic uncertainty relation from the data in the first and second row, with violations occurring in the gray region.}
\label{fig:size}
\end{figure*}

Fig.~\ref{fig:size} shows the mean and variance of the work distribution and the TUR as a function of the system size, with the duration and strength of the protocol fixed at $\tau =1$ and $\gamma=1$, respectively. As the system size increases from $4$ to $130$ spins, the raw quantum computing data in the left column considerably overestimate $\meanwork$ and $\varwork$, without violating the TUR as no point enters the gray region in the lower panel. This behavior is inconsistent with the tensor network data and LRT, predicting that $\meanwork$ and $\varwork$ scale as $|E|^{-1}$, and is a result of device noise. Under extreme depolarizing noise $p(\bts{y},\bts{x}) = 2^{-\nqubits} \langle \bts{x} | \gibbs{0}{\beta} | \bts{x} \rangle$ and thus $\meanwork = \nqubits \tanh(\beta)$ and $\varwork = \nqubits \big( 2 - \tanh(\beta)^2 \big)$, i.e., the momenta of the work distribution are proportional to system size (dashed lines in the left column, see also \supp{E} for a detailed analysis). Nevertheless, raw quantum computing data differ from the extreme depolarizing noise limit and, upon ``SQT'' and ``Ext-SQT'' post-processing, are in qualitative agreement with tensor-network data and LRT (dashed lines in the center and right panels, see \supp{E} for additional comparison). Furthermore, $\meanwork$ and $\varwork$ scale as $|E|^{-1}$, with deviations from LRT occurring for large $\nqubits$ and low $\beta$, due to the subspace approximation. The graph of the function $\nqubits \mapsto \big( \beta^{-2} h(\entropy)(\nqubits), \varwork(\nqubits) \big)$ is a straight line rotating from the vertical axis to the bisector as the inverse temperature decreases, indicating that the TUR is saturated as the inverse temperature vanishes.

\section{Conclusions and Outlook}

The field of quantum thermodynamics is experiencing significant growth, due to its relevance in optimizing design strategies~\cite{auffeves2022_QEI} and in benchmarking and certifying quantum properties in a given process~\cite{onishchenko2022probing}.

Although very recent theoretical results have started shedding light on the role and impact of genuine quantum properties onto the energetic footprint of near-equilibrium processes, in general quantum devices may operate arbitrarily out of equilibrium, for example, when their manipulation requires fast transformations and/or strong coupling.

In this work, we performed an in-depth investigation into the quantum thermodynamic uncertainty relations (TURs) using quantum and classical computers in synergy. 
We proposed and demonstrated a method that uses a quantum computer to sample thermodynamic trajectories of a 2D TFIM and a classical computer to reconstruct the thermodynamics of the system in a subspace defined by such samples.

This method allowed us to perform simulations with up to 130 qubits and circuits of depth 120 (with 87 layers of 2-qubit gates) comprising 8466 operations (of which 4176 are 2-qubit gates), a regime where exact brute-force classical results are not available ~\cite{kim2023evidence}.
In addition, it unambiguously resolved TURs for protocols of different durations and strengths, providing insightful information into the thermodynamics of many-body quantum systems out of equilibrium.

The performance of our method in the present demonstration motivates the study of the (thermo)dynamics of many-body quantum systems in more general and challenging settings, 
with sample-based methods in the framework of quantum-centric supercomputing~\cite{alexeev2024quantum}.
Generalizations of the circuit model presently studied may harbor more intricate behavior, such as a delicate interplay between heat and work. Extensions and adaptations of the sample-based method presented here may be used to study many-body states beyond ground- and low-energy Hamiltonian eigenstates \cite{kanno2023qsci, robledomoreno2024, barison2024quantum}, as well as to probe Landauer's principle ~\cite{landauer1961irreversibility,reeb2014improved,berut2012experimental, Hong2016, Yan2018,aimet2024experimentally} in the many-body and out-of-equilibrium regime. The extent to which the techniques presented here can be extended to more challenging situations, particularly when the initial Hamiltonian is not diagonal in the computational basis or the driving protocol is not cyclical, represents an open problem. Nevertheless, we regard the accuracy of our simulations and the effectiveness of our post-processing techniques as encouraging signals of progress in that direction.

\begin{acknowledgments}
MM thanks Kevin J. Sung for access to the Clifford cluster, Petar Jurcevic, William Kirby, Ewout van der Berg, Yukio Kawashima, and Xuan Wei for useful feedback, and Abhinav Kandala and Maika Takita for access to quantum computers.
MM and GG acknowledge meaningful conversations with Alireza Seif and Mirko Amico.
\end{acknowledgments}

\newpage
\newpage

$ $

\newpage
\newpage

\appendix

\onecolumngrid

\section*{Organization of the appendix}

The appendix is structured as follows.
\begin{itemize}
\item Section~\ref{sec:driving_protocol} provides additional information about the driving protocol, to complement Fig.~\ref{fig:setup} and the discussion around Eq.~\eqref{eq:hamiltonian_main} of the main text, with particular focus on quantum circuits.
\item Section~\ref{eq:work_distribution} provides additional information about the quantum work distribution, with subsections dedicated to: 
\begin{enumerate} 
\item the verification of TURs
\item the post-processing of sampled bitstrings (raw data, QST, and Ext-QST post-processing)
\item the analytic expression for the momenta of the work distribution within linear response theory
\end{enumerate}
\item Section~\ref{sec:appendix_details} provides additional information about the numerical simulations performed in this work, including the qubit layouts used for the simulations in Fig.~\ref{fig:scan}, Fig.~\ref{fig:coup}, and Fig.~\ref{fig:size} of the main text. These simulations, respectively, for different values of protocol duration $\tau$, protocol strength $\gamma$, and system size $\nqubits$, are labeled ``scan'', ``coupling'', and ``size'', respectively in the Appendix.
\item Section~\ref{sec:data} presents additional data, to complement Fig.~\ref{fig:scan}, Fig.~\ref{fig:coup}, and Fig.~\ref{fig:size} of the main text assessing the accuracy of SQT/Ext-SQT error mitigation and the convergence of quantum circuits with number of Trotter steps
\item Section~\ref{sec:appendix_white} complements Fig.~\ref{fig:size} of the main text by
\begin{enumerate}
\item deriving the analytic expression for the momenta of the work distribution assuming an extreme depolarizing channel (also referred to as ``white noise'')
\item comparing the momenta of the work distributions from quantum hardware and an extreme depolarizing channel.
\end{enumerate}
\end{itemize}

\section{Driving Protocol}
\label{sec:driving_protocol}

In this work, we consider a system of $n$ qubits corresponding to the vertices of a graph $(V,E)$ specified in \supp{\ref{sec:appendix_details}}. At the beginning of the driving protocol, the qubits are in equilibrium at inverse temperature $\beta$ under the Hamiltonian
\begin{equation}
\label{eq:h0}
H_0 = - \sum_{q \in V} Z_q \;.
\end{equation}
In the first step of the protocol, the qubits are measured in the Pauli-$\pauli{Z}$ basis, returning a bitstring $\bts{x}$ that follows the distribution
\begin{equation}
\label{eq:initial_bts}
p_{\mathrm{th}}(\bts{x}) = \langle \bts{x} | \gibbs{0}{\beta} | \bts{x} \rangle = \prod_{q \in V} q(x_q)
\;,
\end{equation}
where
\begin{equation}
q(x) = \frac{e^{-\beta E_x}}{\sum_{y=0,1} e^{-\beta E_y}}
\;,\;
E_x = (-1)^{x+1}
\;,
\end{equation}
and collapsing the qubits onto the computational basis state $\ket{\bts{x}} = \otimes_{q \in V} \ket{x_q}$.

In the second step of the protocol, the qubits evolve for a variable time $\tau$ under the action of the time-dependent Hamiltonian $H_t = H_0 + \lambda_t V$, where
\begin{equation}
V = \frac{1}{|E|} \sum_{\edge{p}{r} \in E} X_p X_r
\;,\;
\lambda_t = \gamma \sin\left( \frac{\pi t}{\tau} \right)
\;.
\end{equation}
We approximate the exact time-evolution operator
\begin{equation}
\label{eq:exact_time_evolution}
U_0^\tau = \mathrm{T\exp}\left[ -i \int_0^\tau dt \, H_t \right]
\end{equation}
with a second-order Trotter-Suzuki product formula comprising $\ntrot$ steps,
\begin{equation}
\label{eq:time_evolution}
U_{\tau,\ntrot,\gamma} = \prod_{\ell=0}^{n_T-1} 
e^{-i \frac{\Delta t}{2} H_0}
e^{-i \Delta t w_\ell V}
e^{-i \frac{\Delta t}{2} H_0}
\;,
\end{equation}
where $\Delta t = \frac{\tau}{\ntrot}$, $t_\ell = \left(\frac{1}{2}+\ell \right) \Delta t$, and $w_\ell = \gamma \sin\big( \frac{\pi t}{\tau} \big)$.

In the third step of the protocol, the qubits are measured again in the Pauli-$\pauli{Z}$ basis, returning a bitstring $\bts{y}$ that follows the distribution
\begin{equation}
\label{eq:final_bts}
p(\bts{y}|\bts{x}) = | \braket{\bts{y}}{U_{\tau,\ntrot,\gamma}}{\bts{x}} |^2
\;.
\end{equation}
The protocol is repeated $N$ times, returning a collection of bitstring pairs $\{ (\bts{x}_k, \bts{y}_k) \}_{k=1}^N$.

\subsection{Quantum circuits}
\label{sec:quantum_circuits}

\paragraph{Preparation and initial measurement} 

Consider a Hamiltonian of the form $H_0 = V D V^\dagger$ where $V$ is a known unitary that can be efficiently implemented by a quantum circuit, and $D = \sum_{\bts{x}} D_{\bts{x}} | \bts{x} \rangle \langle \bts{x} |$ a known diagonal operator. The Hamiltonian Eq.~\eqref{eq:h0} is clearly of this form, with $V=I$ and $D_{\bts{x}} = E_{\bts{x}}$.
For such a Hamiltonian, there are at least two simple procedures to sample the thermal states $\gibbs{0}{\beta}$.

(i) For all $k=1\dots N$, sample $\bts{x}_k$ from the distribution $p_{\mathrm{th}}(\bts{x}) \propto e^{-\beta D_{\bts{x}}}$ on a classical computer. Then, form the (up to $N$) quantum circuits $V^\dagger U_{\tau, \ntrot,\gamma} V | \bts{x}_k \rangle$. For each of them, measure all qubits in the Pauli-$\pauli{Z}$ basis, obtaining the bitstring $\bts{y}_k$, distributed according to $p(\bts{y}_k|\bts{x}_k) = | \langle \bts{y}_k | V^\dagger U_{\tau, \ntrot,\gamma} V | \bts{x}_k  \rangle |^2$.

(ii) Form a single circuit, $\mathcal{M} V^\dagger U_{\tau, \ntrot,\gamma} V \mathcal{M} | \Psi \rangle$, where $| \Psi \rangle \propto \sum_{\bts{x}} \sqrt{p_{\mathrm{th}}(\bts{x})} \, |\bts{x}\rangle$ and $\mathcal{M}$ denotes a measurement in the Pauli-$\pauli{Z}$ basis.
Execute this circuit $N$ times, producing a set of $N$ bitstring pairs $\{ (\bts{x}_k, \bts{y}_k) \}_{k=1}^N$ following the desired distribution. 

Approach (i) offers at least two benefits. The sampling of $\bts{x}_k$ is carried out in a noise-free environment (a classical computer), and it is not necessary to prepare the state $|\Psi\rangle$. On the other hand, one has to execute a large number of circuits (up to $N$) using a queue system with a limited (and possibly smaller than $N$) number of circuits per job.

Approach (ii), although it requires the preparation of $|\Psi\rangle$ and relies on noisy mid-circuit measurements, allows one to execute only 1 circuit drawing $N$ samples and therefore makes more efficient use of a queue system.

For the Hamiltonian Eq.~\eqref{eq:h0}, the state $|\Psi\rangle$ can be prepared with a single layer of $\pauli{Y}$ rotations,
\begin{equation}
\label{eq:psibeta}
\ket{ \Psi } = \bigotimes_{q\in V} \Big[ R_{\pauli{Y}}(\theta_\beta) \ket{0} \Big]_q
\;,\;
\cos^2 \left( \frac{\theta_\beta}{2} \right) = q(0)
\;.
\end{equation}
where the angle $\theta_\beta$ is chosen so that $| \overlap{ \bts{x}} {\Psi} |^2 = p_{\mathrm{th}}(\bts{x})$. Therefore, in this study, we elected to pursue approach (ii).
We remark that approach (i) may be preferable in many situations.

\paragraph{Driving protocol}

The quantum circuits that correspond to the operators in Eq.~\eqref{eq:time_evolution} are
\begin{equation}
\label{eq:evo_h0}
e^{-i \frac{\Delta t}{2} H_0}
= 
\prod_{q \in V} e^{i \frac{\Delta t}{2} Z_q}
= 
\bigotimes_{q\in V} R_{\pauli{Z}}\left( \alpha \right)_q
\end{equation}
with $\alpha = -\Delta t$ and
\begin{equation}
\label{eq:evo_vt}
e^{-i \Delta t \, w_\ell V}
=
\prod_{\edge{p}{r} \in E} e^{-i \frac{\Delta t \, w_\ell}{|E|}  X_p X_r}
=
\bigotimes_{\edge{p}{r}  \in E} R_{\pauli{XX}}\left( \beta_\ell \right)_{pr}
\end{equation}
with $\beta_\ell = (2 \Delta t \, w_\ell)/|E|$. The circuit in Eq.~\eqref{eq:evo_h0} is compiled into a single layer of single-qubit $\pauli{Z}$ rotations, and the circuit in Eq.~\eqref{eq:evo_vt}, upon coloring the edges of $(V,E)$ with Vizing's algorithm using $c$ colors \cite{vizing1965critical}, is compiled into $c$ layers of two-qubit $\pauli{XX}$ rotations.
The cost of the quantum circuits run in this work is summarized in Table~\ref{tab:circuits}, and quantitative estimates are reported in \supp{\ref{sec:appendix_details}} along with the specific graphs $(V,E)$ used in this work. For these graphs, $c=3$.

\begin{table}
\begin{tabular}{cccccc}
\hline\hline
 & step & depth & type & number \\
\hline
 & \eqref{eq:psibeta} & 1 & $R_\pauli{Y}$ & $n$ \\ 
 & meas. $\bts{x}$ & 1 & $\pauli{Z}$ meas. & $n$ \\ 
& \eqref{eq:evo_h0} & $(\ntrot+1)$ & $R_\pauli{Z}$ & $n$ \\ 
& \eqref{eq:evo_vt} & $\ntrot c$ & $R_\pauli{XX}$ & $|E|/c$ \\ 
& meas. $\bts{y}$ & 1 & $\pauli{Z}$ meas. & $n$ \\ 
\hline\hline
\end{tabular}
\caption{Cost of the quantum circuits ran in this work, expressed by the number and type (column 4 and 3, respectively) of the quantum operations occurring in the steps of the driving protocol (column 1), and the number of layers (column 2) in which these operations are arranged.
}
\label{tab:circuits}
\end{table}

\section{Work distribution}
\label{eq:work_distribution}

\subsection{Verifying thermodynamic uncertainty relations}

Let us consider a system, initially prepared in a thermal Gibbs state $\gibbs{0}{\beta}$ at inverse temperature $\beta$ (e.g. by equilibration with an external thermal reservoir that is detached before the start of the drive protocol), where we denote $\gibbs{t}{\beta} = e^{-\beta H_t}/Z_t$, with $Z_t = \mathrm{Tr}[e^{-\beta H_t}]$ being the partition function. 
The system is then driven out of equilibrium by means of an external time-dependent driving protocol $t\mapsto\lambda_t$ acting, for example, on some parameter of the system Hamiltonian. As a result of this operation, the system after a time $\tau$ is in the state $\rho_\tau = U_\tau \gibbs{0}{\beta} U^\dagger_\tau$, with $U_\tau$ as in Eq.~\eqref{eq:exact_time_evolution}, which is in general different from the final equilibrium state $\gibbs{\tau}{\beta}$.
The amount of irreversibility associated with this transformation equals the \textit{dissipated work}, namely $\Sigma = \beta(\meanwork - \Delta F) \geq 0$, where $\meanwork$ denotes the total work done on the system by the external agent through the driving protocol and $\Delta F = \ln(Z_\tau/Z_0)$ indicates the free-energy difference (i.e. the average minimum work) ~\cite{Callen1951,jarzynski1997nonequilibrium}. It is worth pointing out that $\Sigma = D(\rho_\tau||\gibbs{\tau}{\beta})$, with $D(\rho||\sigma) = \mathrm{Tr}\left[\rho\ln(\rho/\sigma)\right]$ being the quantum relative entropy.
No irreversible entropy production is generated if the process is adiabatic, i.e. $\rho_t = \gibbs{t}{\beta}$ $\forall t \in [0,\tau]$, in which case $\meanwork = \Delta F$~\cite{Brandner2020}.
We emphasize that this definition of the entropy production $\Sigma$ coincides with the most familiar expression by Clausius if the environment is not detached after the initial state preparation and is kept in contact with the system throughout the external driving. In this expression $\Sigma = \Delta S - \beta \meanheat$, with $\meanheat$ being the heat absorbed from by the environment and $\Delta S$ the change of the system's entropy~\cite{mohammady2020energetic}. The duality of these two formulations crucially relies on the assumption of initial equilibrium, which allows to connect changes in the free energy and in the entropy through $\Delta F = \meaninten - \beta^{-1}\Delta S $, with $ \meaninten$ being the change in the internal energy of the system. Using the first law of thermodynamics $\meaninten = \meanwork + \meanheat$, one can then easily show that
\begin{equation}
\Sigma = \Delta S - \beta \meanheat = \Delta S - \beta \left(\meaninten - \meanwork\right) = \beta\meanwork - \beta\left( \meaninten - \beta^{-1}\Delta S \right) = \beta \left(\meanwork-\Delta F\right).
\label{eq:equivalence}
\end{equation}
In our setup we will therefore identify, without loss of generality, the entropy production with the dissipated work and will focus on the latter quantity. 
Moreover, since we will perform a cyclical driving protocol, we will have that $\Delta F = 0$ and therefore the entropy production will become proportional to the work done on the system along the protocol $\Sigma = \beta \meanwork$.
According to the theory of classical stochastic Thermodynamics, this quantity represents a random variable distributed according to some probability distribution that depends on the driving protocol as well as on the temperature and other relevant Hamiltonian parameters.
In the quantum regime, the work also crucially depends on the measurement scheme chosen to access it; a standard choice is through the so-called two-point measurement (TPM) scheme, which consists in measuring the system's energy at the beginning and at the end of the protocol, and identifying the work with the difference in the two measurement outcomes~\cite{Talkner2007b}. In our setup this is fully justified by the fact that the system is isolated during the computation from an external environment, so that any change of its internal energy can only be due to the work performed on the system. 
Explicitly, if one denotes with $E_0$ and $E_\tau$ the two outcomes of the system's energy projective measurement, the the work defined as $\mathcal{W} = E_\tau - E_0$ becomes a classical stochastic variable distributed according to the Born rule
\begin{equation}\label{eq:tpmprob}
    p(\mathcal{W}) = \sum_{E_\tau - E_0 = \mathcal{W}} \bra{E_0}\gibbs{0}{\beta}\ket{E_0} \,\cdot\,\left|\langle E_\tau | U^\tau_0| E_0\rangle\right|^2,
\end{equation}
where $\ket{E_t}$ denotes the eigenvector associated to the outcome $E_t$ (assuming for simplicity that there is no degeneracy), i.e. $H_t = \sum_{E_t} E_t \ket{E_t}\bra{E_t}$.
Knowledge of the probability distribution Eq.~\eqref{eq:tpmprob} or of its Fourier transform, i.e. the so-called generating function, allows to compute the various cumulants of the work, e.g. the average and the variance.

The typical form of TURs involves the ``signal-to-noise ratio'' (SNR) of the work $\snr = \frac{ \varwork }{ \meanwork^2 }$ and the entropy production $\entropy = \beta \, \meanwork$ of the quantum work distribution, and takes the form
\begin{equation}
\label{eq:tur}
\snr \geq f(\entropy) 
\;,
\end{equation}
where $f(x) = \sinh^{-2}\left( g(x/2) \right)$ and $g^{-1}(x) = x \tanh(x)$. The mean and variance of the quantum work distribution scale as $O(\gamma^2)$ in the linear-response regime, and thus the SNR becomes difficult to resolve numerically.
For this reason, and for ease of graphical representation, in this study we elected to verify the TUR Eq.~\eqref{eq:tur} by considering a lower bound for the work variance defined by the entropy production. More specifically, multiplying both sides of Eq.~\eqref{eq:tur} by $\meanwork^2$ and observing that $\meanwork = \beta^{-1} \entropy$, one readily obtains
\begin{equation}
\varwork \geq \beta^{-2} \entropy^2 f(\entropy) \equiv \beta^{-2} h(\entropy)
\;,
\end{equation}
where we defined $h(x) = x^2 f(x)$.

\subsection{Post-processing of sampled bitstrings}

The bitstring pairs $\{ (\bts{y}_k,\bts{x}_k) \}_{k=1}^N$ accumulated in the protocol are post-processed in three different ways. The post-processing operations are preceded by post-selection based on conservation of parity: as the time-dependent Hamiltonian $H_t$ commutes with the parity operator $P= \prod_{q \in V} Z_q$, the elements in each pair $(\bts{y}_k,\bts{x}_k)$ should have identical parity, i.e., $\sum_q (x_k)_q = \sum_q (y_k)_q$ mod 2. When this condition is not met, the corresponding pair is discarded, resulting in a lower sample efficiency.

Symmetry breaking is a significant limitation, e.g., in electronic structure calculations where, due to quantum noise, the sampling efficiency can decrease to $0$ as circuits become larger~\cite{robledomoreno2024}. In the present case, however, the sampling efficiency is always above $50 \%$, and while it may be increased with a suitable generalization of configuration recovery~\cite{robledomoreno2024}, in this work we elected to use post-selection based on conservation of parity.

\paragraph{Post-processing of raw data} First, one can simply compute the sample average and variance of the quantum work distribution as
\begin{equation}
\label{eq:raw_estimator}
\begin{split}
\meanwork &= \frac{1}{N} \sum_{k=1}^N \Big( E_{\bts{y}_k} - E_{\bts{x}_k} \Big) \;, \\
\varwork  &= \frac{1}{N-1} \sum_{k=1}^N \Big( E_{\bts{y}_k} - E_{\bts{x}_k} - \meanwork \Big)^2 \;, \\
\end{split}
\end{equation}
where $E_{\bts{x}} = \braket{\bts{x}}{H_0}{\bts{x}} = \sum_{q \in V} (-1)^{x_q+1}$. 

Data post-processed in this way are labeled ``raw'' in the main text.

\paragraph{SQT post-processing} On a noisy quantum device, computational basis states $(\bts{y}_k,\bts{x}_k)$ are sampled from a distribution that differs from the ideal one, $p(\bts{y},\bts{x}) = p(\bts{y}|\bts{x}) \, p(\bts{x})$ with marginal and conditional distributions defined in Eq.~\eqref{eq:initial_bts} and Eq.~\eqref{eq:final_bts} respectively. As a result, the raw estimators in Eq.~\eqref{eq:raw_estimator} are biased. Motivated by sample-based quantum diagonalization, we introduce a technique to mitigate such a bias. First, we define the following set of computational basis states,
\begin{equation}
S = \{ \bts{y}_k \}_{k=1}^N \cup \{ \bts{x}_k \}_{k=1}^N \equiv \{ \bts{z}_l \}_{l=1}^M
\;.
\end{equation}
We then calculate, using a classical computer, the length-$M$ vector (note that $M \leq 2N$)
\begin{equation}
\tilde{E}_m = E_{\bts{z}_m} \;,
\end{equation}
and the matrix (note that this matrix is sparse, with at most $|E|$ non-zero entries per row)
\begin{equation}
\label{eq:projected_v}
\tilde{V}_{mn} = \braket{\bts{z}_m}{V}{\bts{z}_n}
\;,
\end{equation}
and use them to define the probability distribution
\begin{equation}
\tilde{p}_{mn} = T_{mn} \frac{ e^{-\beta \tilde{E}_n} }{\tilde{Z}}
\end{equation}
with $\tilde{Z} = \sum_n e^{-\beta \tilde{E}_n}$,
\begin{equation}
\label{eq:compute_T}
T_{mn} = \braket{\bts{z}_m}{\prod_{\ell=0}^{\ntrot-1} 
e^{-i \frac{\Delta t}{2} \tilde{D}} 
e^{-i \Delta t w_\ell \tilde{V}} 
e^{-i \frac{\Delta t}{2} \tilde{D}}}{\bts{z}_n}
\;,
\end{equation}
and $\tilde{D} = \mbox{diag}(\tilde{E})$. We then use the computed probability distribution to define the  following estimators,
\begin{equation}
\label{eq:qcs_estimator}
\begin{split}
\meanwork &= \sum_{mn=1}^M \tilde{p}_{mn} \Big( E_{\bts{y}_m} - E_{\bts{x}_n} \Big) \;, \\
\varwork  &= \sum_{mn=1}^M \tilde{p}_{mn} \Big( E_{\bts{y}_m} - E_{\bts{x}_n} - \meanwork \Big)^2 \;. \\
\end{split}
\end{equation}
The most expensive part of the procedure is the calculation of Eq.~\eqref{eq:compute_T}, that we carry out using the \library{expm\_multiply} of \library{scipy}'s \library{sparse.linalg} library, at cost at cost $O(M^2)$.
Data post-processed in this way are labeled ``SQT'' in the main text. Eq.~\eqref{eq:qcs_estimator} corresponds to projecting the time-dependent Hamiltonian $H_t$ in the subspace spanned by bitstrings in $S$ through the operator $P = \sum_{m=1}^M \ket{\bts{z}_m} \bra{\bts{z}_m}$, and classically computing the Gibbs distribution defined by $PH_0P$ and the matrix elements of the time-evolution operator defined by $PH_tP$.
The estimator Eq.~\eqref{eq:qcs_estimator} is biased by the projection, to an extent that depends on multiple factors, particularly: the size of $S$, the strength of the noise affecting samples drawn from a specific quantum computer, the concentration properties of the wavefunctions $U_{\tau,\ntrot,\gamma} | \bts{x} \rangle$, and the inverse temperature $\beta$. A fundamental limitation of the SQT post-processing is that, in the limit of low inverse temperature, $\gibbs{0}{\beta} \to I/2^{\nqubits}$ resulting in the breakdown of the sparsity approximation.

\begin{figure}[h!]
\includegraphics[width=0.4\columnwidth]{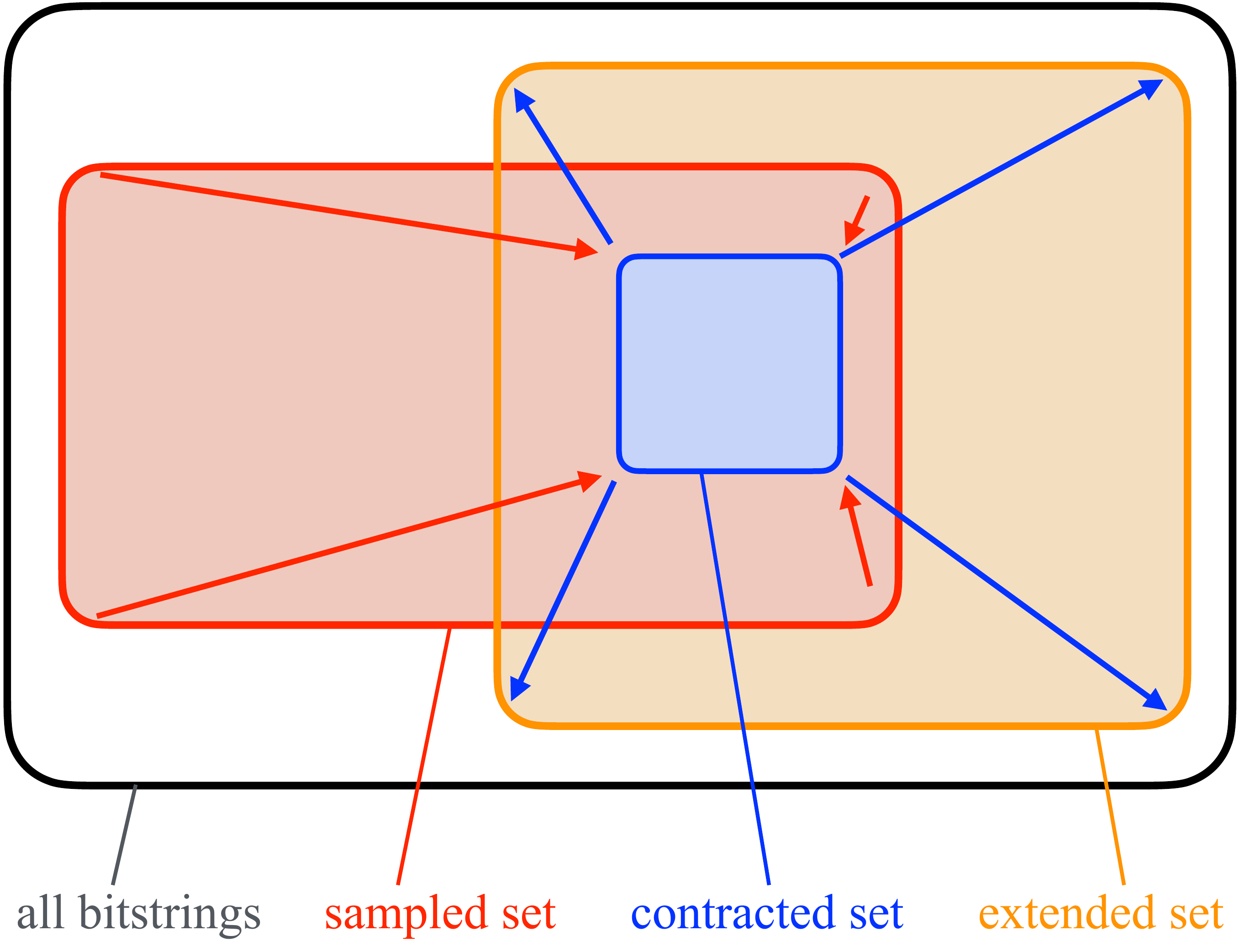}
\caption{\figuretitle{Illustration of the extended SQD-based post-processing.} The set of sampled configurations (red, $S$ in the text) is first pruned to remove configurations with low probability over the Gibbs state, producing a contracted set (blue), then extended to include all configurations connected to those in the contracted set (though not in the sampled set) by the action of the interaction Hamiltonian $V$, resulting in an extended set (orange, $S_{\mathrm{ext}}$ in the text).}
\label{fig:extended}
\end{figure}

\paragraph{Ext-SQT post-processing} An important limitation of the ``SQT'' estimator in Eq.~\eqref{eq:qcs_estimator} is that two configurations $\bts{z}_m,\bts{z}_n \in S$ may not be connected by the interaction Hamiltonian $V$, i.e. $\tilde{V}_{mn}=0$ for several entries of the matrix $\tilde{V}$ in Eq.~\eqref{eq:projected_v}. When this happens, $\tilde{p}_{mn}$ is a diagonally dominated matrix, leading to many small or vanishing contributions in Eq.~\eqref{eq:qcs_estimator}.
To remedy this limitation, we introduce an extension of the SQD-based post-processing described in the previous paragraph, motivated by the extension of SQD to excited-state electronic structure calculations \cite{barison2024quantum}, wherein the set $S$ is (see Fig.~\ref{fig:extended}) first pruned to remove configurations with low probability over the equilibrium state $\gibbs{0}{\beta}$, yielding a contracted set of configurations $C$, then extended as follows:
\begin{equation}
S_{\mathrm{ext}} = \bigcup_{\bts{x} \in C} \mathrm{N}[\bts{x}] \;,
\end{equation}
where $\mathrm{N}[\bts{x}] = \{ \bts{z} \in \{0,1\}^{\nqubits} : \braket{\bts{z}}{V}{\bts{x}} \neq 0\}$.
For a given bitstring $\bts{x}$, the set $\mathrm{N}[\bts{x}]$ can be determined at polynomial cost, because it contains bistrings of the form $\bts{z} = (\bts{x} + \bts{e}_p + \bts{e}_r)$ mod 2, where $(\bts{e}_p)_q = \delta_{pq}$ and $\edge{p}{r} \in E$. Therefore, $|\mathrm{N}[\bts{x}]| = |E|$ and $|S_{\mathrm{ext}}| \leq |C| |E| \leq M |E| \leq 2N |E|$.
The operations described in the previous paragraph are carried out for the set $S_{\mathrm{ext}}$ without any modifications, and data post-processed in this way are labeled ``Ext-SQT'' in the main text.

The extension from $S$ to $S_{\mathrm{ext}}$ ensures that any element of $S_{\mathrm{ext}}$ is connected to one or more elements of the same set by the interaction Hamiltonian $V$. This mitigates the diagonal domination of $\tilde{p}$ and yields a more accurate and efficient estimator for $\meanwork$ and $\varwork$.

\subsection{Linear response theory analysis}

Let us consider a quantum system, unitarily driven out of equilibrium through a time-dependent Hamiltonian $H_t =H_0+\lambda_t V$ acting over a finite time interval $0 \leq t \leq \tau$, where $t\mapsto \lambda_t$ is the dimensionless driving protocol, and the operator $V$ (with normalization $\|V\|=1$) is treated as an external perturbation that is turned on at time $t=0$ and off at time $t=\tau$ (i.e.,~$\lambda_0=\lambda_\tau=0$). 
The system, still initially prepared in the Gibbs state $\gibbs{0}{\beta}$ at inverse temperature $\beta$ (e.g. by equilibration with an external thermal reservoir that is detached before the start of the drive protocol), is brought out of equilibrium by means of an external Hamiltonian perturbation $\rho_\tau = U^\tau_0 \gibbs{0}{\beta} \left( U^\tau_0 \right)^\dagger$, with $U^\tau_0$ as in Eq.~\eqref{eq:exact_time_evolution}.

If the perturbation is weak, i.e. $|\lambda_t|\ll 1$ for all $0 \leq t \leq \tau$, the state is expected to remain close to the initial equilibrium state at all times \cite{bruus2004many},
\begin{equation}
\label{eq:kubo}
\rho_t
= 
\gibbs{0}{\beta}
- i \int^t_0 dt' \, \lambda_{t'}[V(t-t'),\gibbs{0}{\beta}]
\;,
\end{equation}
with $A(t) = e^{i H_0 t} A e^{-iH_0 t}$ indicating the interaction picture. This regime, known as linear response theory (LRT), has retained a central role in the study of systems close to equilibrium~\cite{marconi2008fluctuation} since its original formulation by Kubo~\cite{Kubo1957}, leading to several important results in quantum transport \cite{fu1993quantum}, many-body quantum physics \cite{suzuki1968dynamics}, quantum field theory \cite{calzetta2009nonequilibrium} and optimal control \cite{Topical,Deffner2015c,Deffner2018,Deffner2018a}. 
Under this approximation, it is known that the first-order correction to the average dissipated work is given in terms of the two-time integral ~\cite{andrieux2008,Deffner2015c,Deffner2018}
\begin{align}\label{eq:av}
\beta \, \meanwork = \frac{1}{2}\int^\tau_0 dt \int^\tau_0 dt'  \ \Psi_0(t-t')\dot{\lambda}_t \dot{\lambda}_{t'}.
\end{align}
Here $t\mapsto \Psi_0(t)$ denotes the central object in LRT, known as the Kubo relaxation function~\cite{Kubo1957}
\begin{align}\label{eq:kubocov}
\Psi_0(t) = \beta\int^\beta_0 ds \ \langle V(-i\hbar s)V(t)\rangle-\beta^2 \langle V \rangle^2,
\end{align}
with $\langle .\rangle$ denoting the average with respect to the thermal state $\gibbs{0}{\beta}$. 
Recently, this framework has been significantly pushed forward in Ref.~\cite{guarnieri2024generalized} by the discovery of closed analytic expressions for the linear corrections to all higher statistical cumulants of the work distribution $\moment{k}$ beyond the mean value $\meanwork = \moment{1}$, e.g. the variance $\varwork \equiv \moment{2}$, which can be found by setting $k=2$ in the general formula
\begin{equation}
\label{eq:cumulants}
\beta^k \moment{k}
=
\int_{-\infty}^{\infty}
\frac{d\omega}{\sqrt{2\pi}}
\tilde{\Psi}_0(\omega)
\gamma_k(\omega)
\bigg|\int^\tau_0 dt \ \dot{\lambda}_t e^{i\omega t}\bigg|^2
\;,
\end{equation}
where 
\begin{align}
\begin{aligned}
\gamma_k(\omega)= 
\left\{
\begin{aligned}
& \frac{1}{2}(\beta\omega)^{k-1}\text{Coth}(\beta\omega/2) && \text{if $k$ even} \\
& \frac{1}{2}(\beta\omega)^{k-1} && \text{if $k$ odd}
\end{aligned}
\right.
\end{aligned} 
\end{align}
These theoretical findings have not yet been applied to real experiments. In this work, we apply them to the TFIM and show that they provide a persuasive framework to interpret experimental values, as well as useful insights into the underlying physics of the computational process.

To this end, let us compute the Kubo relaxation function for the TFIM and use it to extract the momenta of the quantum work distribution. First, we have that
\begin{equation}
\begin{split}
V(-is) &= \sum_{\edge{p}{r} \in E} \frac{1}{|E|} \tilde X_p(s) \tilde X_r(s) \;,\; 
V(t) = \sum_{\edge{p}{r} \in E} \frac{1}{|E|} X_p(t) X_r(t) \\
\end{split}
\end{equation}
with
\begin{equation}
\begin{split}
\tilde X(s) &= e^{-sZ} X e^{sZ} = \ttmat{0}{e^{-2s}}{e^{2s}}{0} \;,\; 
X(t) = e^{-itZ} X e^{itZ} = \ttmat{0}{e^{-2it}}{e^{2it}}{0} \;.
\end{split}
\end{equation}
Therefore,
\begin{equation}
\begin{split}
\langle V(-is) V(t) \rangle 
&= \sum_{\substack{\edge{p}{r} \in E \\ \edge{q}{f} \in E}} 
\frac{\langle \tilde X_p(s) \tilde X_r(s) X_q(t) X_f(t) \rangle}{|E|^2} 
= \sum_{\edge{p}{r} \in E} \frac{\langle \tilde X_p(s) X_p(t) \rangle \langle \tilde X_r(s) X_r(t) \rangle}{|E|^2} 
= \frac{h(s,t)^2}{|E|} \;,
\end{split}
\end{equation}
having observed that the trace $\langle \cdot \rangle$ is zero unless $\edge{p}{r}=\edge{q}{f}$ (because in the Pauli-$\pauli{Z}$ basis $\tilde X(s)$ and $X(t)$ are off-diagonal and the Gibbs state is diagonal) and defined
\begin{equation}
\begin{split}
h(s,t) &= \langle \tilde X(s) X(t) \rangle  =
\frac{e^\beta e^{-2s} e^{2it} + e^{-\beta} e^{2s} e^{-2it} }{e^\beta + e^{-\beta}}
\;.
\end{split}
\end{equation}
Recalling that $\langle V \rangle = 0$ (since in the Pauli-$\pauli{Z}$ basis the operators $H_0$ and $\gibbs{0}{\beta}$ are diagonal, whereas $V$ is off-diagonal), we obtain the following expression for the Kubo relaxation function 
\begin{equation}
\begin{split}
\Psi_0(t) 
&= \frac{\beta}{|E|} \int_0^\beta ds \, h(s,t)^2 
= \frac{\beta}{2 |E|} \left[ \tanh(\beta) \cos(4t) + \beta \mathrm{sech}^2(\beta) \right] \;.
\end{split}
\end{equation}
The Fourier transform of $\Psi_0(t)$ is therefore
\begin{equation}
\label{eq:psi_tilde_zero}
\begin{split}
\tilde \Psi_0(\omega) = \frac{\beta}{2 |E|} &\left[ \tanh(\beta) \, \sqrt{2\pi}  \, \frac{\delta(\omega-4) + \delta(\omega+4)}{2} \right. 
+ \left. \beta \mathrm{sech}^2(\beta) \, \sqrt{2\pi} \, \delta(\omega) \right]
\end{split}
\end{equation}
having recalled that the Fourier transform of $\cos(\omega_0 t)$ is $\sqrt{2\pi} \frac{ \delta(\omega-\omega_0) + \delta(\omega+\omega_0) }{2}$.

To evaluate the momenta of the work distribution, let us first introduce the function
\begin{equation}
\begin{split}
\label{eq:ell}
\left| \int_0^\tau dt \dot \lambda_t e^{i\omega t} \right|^2 
= 
\gamma^2 g(\omega\tau) 
\;,\;
g(\omega\tau) =
2 \pi^2 (\omega\tau)^2 \frac{1 + \cos(\omega\tau)}{(\pi^2 - (\omega\tau)^2)^2}
\;,
\end{split}
\end{equation}
and insert Eq.\eqref{eq:psi_tilde_zero} and \eqref{eq:ell} in Eq.~\eqref{eq:cumulants}, obtaining
\begin{equation}
\begin{split}
\beta^k \moment{k} = \frac{\beta \gamma^2}{2 |E|} &\left[ \tanh(\beta) \frac{\gamma_k(4\beta) g(4\tau) + \gamma_k(-4\beta) g(-4\tau)}{2} \right. 
+ \left. \beta \mathrm{sech}^2(\beta) \gamma_k(0) g(0) \right] 
\;.
\end{split}
\end{equation}
Observing that (i) $g(0)=0$ and $\gamma_k(0)$ is finite, and that (ii) $g(x)$ and 
$\gamma_k(x)$ are even functions, we obtain
\begin{equation}
\moment{k} = \frac{\gamma^2}{|E|} \left[ \frac{\beta^{1-k}}{2} \tanh(\beta) \gamma_k(4\beta) \right] g(4\tau) \;.
\end{equation}
In this result, we recognize a product of four terms, 
(i) the square of the drive protocol strength, $\gamma^2$
(ii) the inverse of the system size, $|E|^{-1}$
(iii) a term that depends exclusively on $\beta$, between square brackets, and 
(iv) a term that depends exclusively on $\tau$, $g(4\tau)$.

The mean and variance of the quantum work distribution are given by
\begin{equation}
\begin{split}
\meanwork &= \frac{\gamma^2}{|E|} \frac{\tanh(\beta)}{4} g(4\tau) 
\;,\;
\varwork  = \frac{\gamma^2}{|E|} \left[ \tanh(\beta) \coth(2\beta) \right] g(4\tau) \;.
\end{split}
\end{equation}
These expressions indicate that:
\begin{enumerate}
\item momenta of the work distribution are inversely proportional to the system size, due to the factor $|E|^{-1}$, which is in agreement with Fig.~\ref{fig:scan} and Fig.~\ref{fig:size} of the main text.
\item mean and variance of the work are maximized, as a function of $\tau$, for $\tau^* = \mathrm{argmax}_\tau g(\tau) \simeq 1.07383$, which is in agreement with Fig.~\ref{fig:scan} of the main text.
\item the lower bound for the work variance can be approximated (in the LRT regime of small $\gamma$) with $\beta^{-2} h(\Sigma) \simeq \beta^{-2} 2 \Sigma = 2 \beta^{-1} \meanwork$, and therefore
\begin{equation}
\begin{split}
\varwork 
&= \frac{ \varwork }{ \meanwork } \meanwork 
= 4 \coth(2\beta) \meanwork 
\simeq 4 \coth(2\beta) \frac{ \beta^{-2} h(\Sigma) }{ 2 \beta^{-1} } 
= 2 \beta \coth(2\beta) \left[ \beta^{-2} h(\Sigma) \right]
\;.
\end{split}
\end{equation}
This equality implies that the graph of the function $\tau \mapsto \big( \beta^{-2} h(\entropy)(\tau), \varwork(\tau) \big)$ is a straight line with a slope
\begin{equation}
m = 2 \beta \coth(2\beta) 
\end{equation}
that, at low inverse temperature $\beta \to 0$, converges to $1$ (bisector) and, at high inverse temperature $\beta \to \infty$, diverges as $\beta$ (vertical axis), which is in agreement with Fig.~\ref{fig:scan} of the main text.
\end{enumerate}

\begin{figure}[b!]
\includegraphics[width=\textwidth]{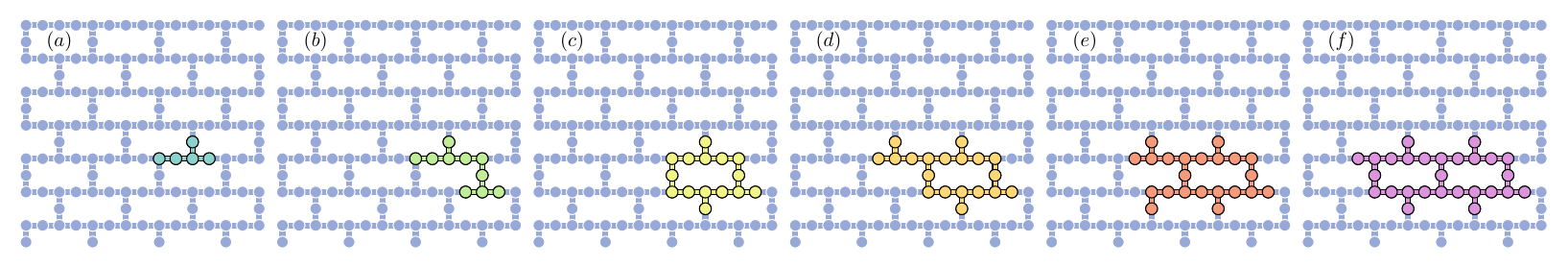}
\caption{\figuretitle{Layouts of the ``scan'' simulations}. These use 5, 10, 15, 19, 22, and 27 qubits (left to right, panels ($a$)-($f$), qubits marked teal, green, yellow, orange, red, and purple) on \device{torino}.}
\label{fig:scan_layouts}
\end{figure}

\begin{figure}[b!]
\includegraphics[width=0.4\columnwidth]{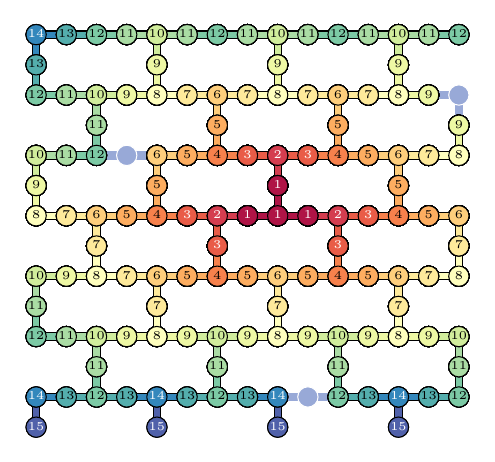}
\caption{\figuretitle{Layouts of the ``size'' simulations}. These are carried out on 15 groups of qubits (numerical indices in white/black) with size ranging from 4 to 130 qubits (colors ranging from red to blue) on \device{torino}. Qubit layouts are a sequence of increasing sets marked with colors from red (layout 1, with 4 qubits) to dark blue (layout 15, with 130 qubits). Qubits marked with an index $k=1\dots 15$ belong to layouts 1 to $k$.}
\label{fig:size_layouts}
\end{figure}

\section{Details of numerical simulations}
\label{sec:appendix_details}

We simulated thermodynamic uncertainty relations of a TFIM on a graph $(V,E)$ where vertices $V$ correspond to a group of qubits on a device with heavy-hex connectivity, and edges $E$ correspond to pairs of qubits in $V$ that are physically connected (i.e., nearest-neighbors in the topology of the device). We carried out three simulations on \device{torino}:
\begin{enumerate}
\item over values of $\tau$ (``scan''), with $\gamma=1$, $\ntrot=4,6,8,10$ and $\nqubits=5,10,15,19,22,27$, using the layouts in Fig.~\ref{fig:scan_layouts}
\item over values of $\gamma$ (``coupling''), with $\tau=1$, $\ntrot = 1 \dots 8$ and $\nqubits=15$, using the layout of Fig.~\ref{fig:scan_layouts}c
\item over values of $\nqubits$ (``size''), with $\tau=1$, $\ntrot = 1 \dots \mathrm{diam}(V,E)$ and $\gamma=1$, using the layouts in Fig.~\ref{fig:size_layouts}
\end{enumerate}
These layouts define the graphs $(V,E)$ introduced in Sec.~\ref{sec:quantum_circuits}, where quantum circuits are described.
For the largest simulation, $(V,E)$ has $\nqubits=|V|=130$ qubits and $|E| = 144$ edges (layout marked ``15'' in Fig.~\ref{fig:size_layouts}), that Vizing's algorithm allows to color using $c=3$ colors. We ran quantum circuits with up to $\ntrot=29$ Trotter-Suzuki steps (because $\mathrm{diam}(V,E)=29$ for the largest layout in Fig.~\ref{fig:size_layouts}), resulting in:
\begin{itemize}
\item $\nqubits=130$ $R_{\pauli{Y}}$ gates, 1 layer
\item $\nqubits=130$ Pauli-$\pauli{Z}$ mid-circuit measurements, 1 layer
\item $\nqubits (\ntrot + 1) = 3900$ $R_{\pauli{Z}}$ gates, $(\ntrot+1)=30$ layers
\item $|E| \ntrot = 4176$ $R_{\pauli{XX}}$ gates, $c \ntrot = 87$ layers
\item $\nqubits=130$ Pauli-$\pauli{Z}$ measurements, 1 layer
\end{itemize}
and thus in a depth of $1+1+30+87+1= 120$ (with $87$ layers of 2-qubit gates) and a total of $130+130+3900+4176+130=8466$ quantum operations (of which 3900 are $\pauli{Z}$ rotations, implemented virtually on superconducting quantum computers, and 4176 are 2-qubit gates).

For each circuit, we collected $20000$ measurement outcomes (also called ``shots'' in the quantum computing literature). Our simulations used readout error mitigation (ROEM)~\cite{nation2021scalable} to mitigate errors arising from qubit measurement, and dynamical decoupling (DD)~\cite{viola1998dynamical,kofman2001universal,biercuk2009optimized} to mitigate errors arising from quantum gates. We used the implementation of ROEM and DD available in \library{Qiskit}'s \library{Runtime} library~\cite{Qiskit}, through the \library{SamplerV1} primitive. Our DD protocol applies two $\pauli{X}$ pulses (as in Ramsey echo experiments) to idle qubits.

We performed tensor network simulations~\cite{schollwock2011density} using \library{Qiskit}'s matrix product state (MPS) simulation backend and the time-evolving block decimation method (TEBD) as implemented in \library{Qiskit}'s \library{Aer} 0.15.0 library. Due to resource constraints, the MPS simulation backend supports up to 32 qubits, as described in the function $\mathsf{backend.configuration().\_qubits}$.

Post-processing involved projections on subspaces of dimension up to $32000$, parallelized over values of $\tau$, $\gamma$, $\nqubits$, $\ntrot$, and $\beta$ using the MPI4PY library, and carried out using up to 120 cores on a bare metal node consisting of four sockets, each with an Intel Xeon Platinum 8260 (2.40GHz) processor, using in-house software.

\section{Additional data}
\label{sec:data}

\subsection{Simulations over values of $\tau$ (``scan'')}

\begin{figure}[b!]
\includegraphics[width=\textwidth]{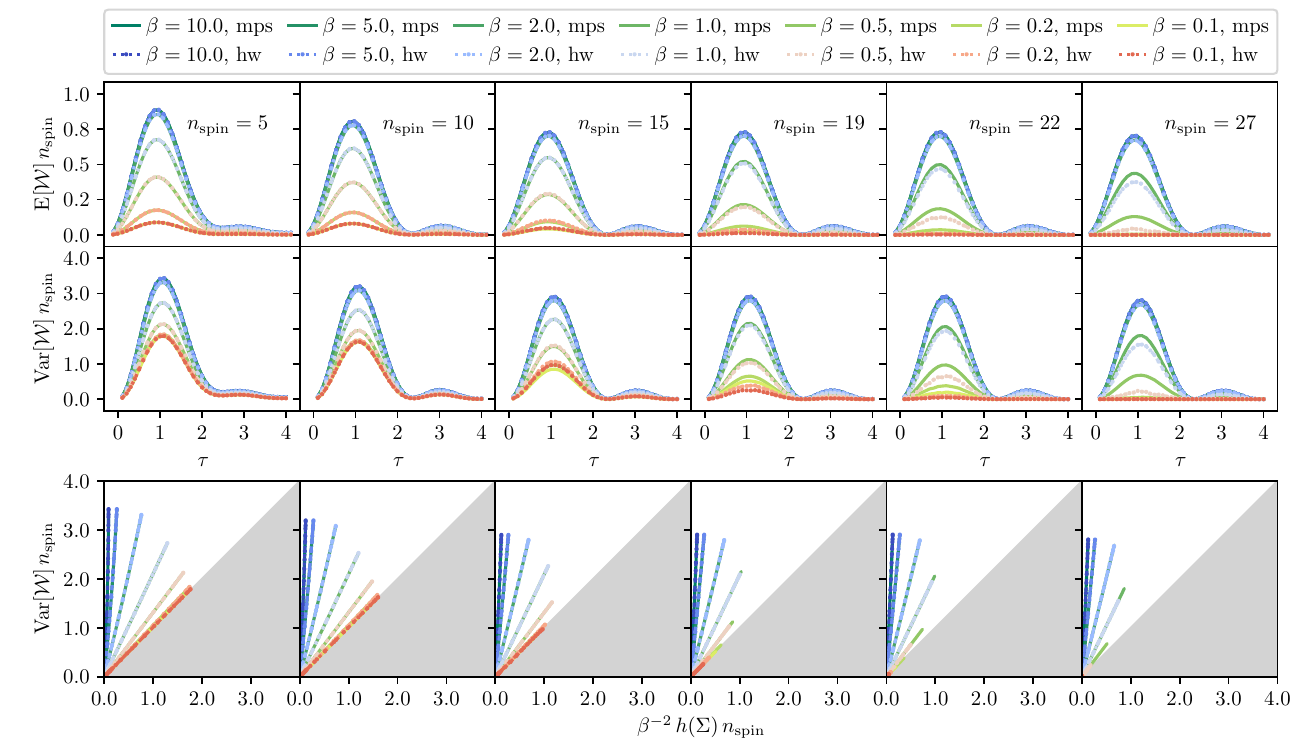}
\caption{\figuretitle{``scan'' simulations.} First and second row: mean and variance of the quantum work distribution as a function of protocol duration $\tau$, (left to right), computed on \device{torino} (dotted curves, ``hw'') and with matrix product states (solid curves, ``mps'') for $\nqubits = 5$ to 27 qubits and inverse temperatures from $\beta=10.0$ to $0.1$ (blue to red for ``hw'' and dark to light geen for ``mps''). Third row: TUR from the data in the first two rows, with violations occurring in the light gray region.
}
\label{fig:figure_2_supplement}
\end{figure}

In Fig.~\ref{fig:figure_2_supplement}, we show all the simulations scanning values of $\tau$, of which those in Fig.~\ref{fig:scan} are a subset.

\begin{figure}[h!]
\includegraphics[width=\textwidth]{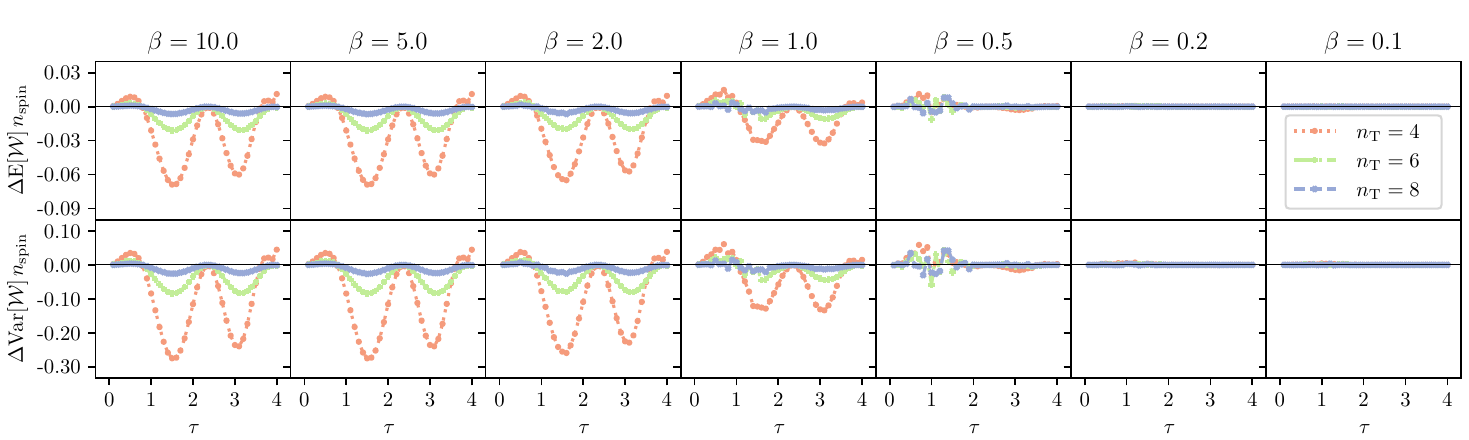}
\caption{\figuretitle{Trotter error in the ``scan'' simulations.} First row: deviation $\Delta \meanwork = \meanwork(\ntrot) - \meanwork(\ntrot=10)$ in the mean of the quantum work distribution computed with $\ntrot = 4,6,8$ and $\ntrot = 10$ Trotter-Suzuki steps, for $\nqubits = 27$ qubits and inverse temperatures $\beta = 10.0, 5.0, 2.0, 1.0, 0.5, 0.2, 0.1$ (left to right), from mitigated quantum hardware data. Second row: deviation $\Delta \varwork$ in the variance of the quantum work distribution under the same conditions as in the first row. 
}
\label{fig:trotter_scan}
\end{figure}

In Fig.~\ref{fig:trotter_scan}, we assess the impact of the Trotter approximation on the results in Fig.~\ref{fig:figure_2_supplement}.
Deviations are more pronounced at higher inverse temperatures (left) and appear less continuous at lower inverse temperatures (right) due to finite sampling effects.

\begin{figure}[h!]
\includegraphics[width=\textwidth]{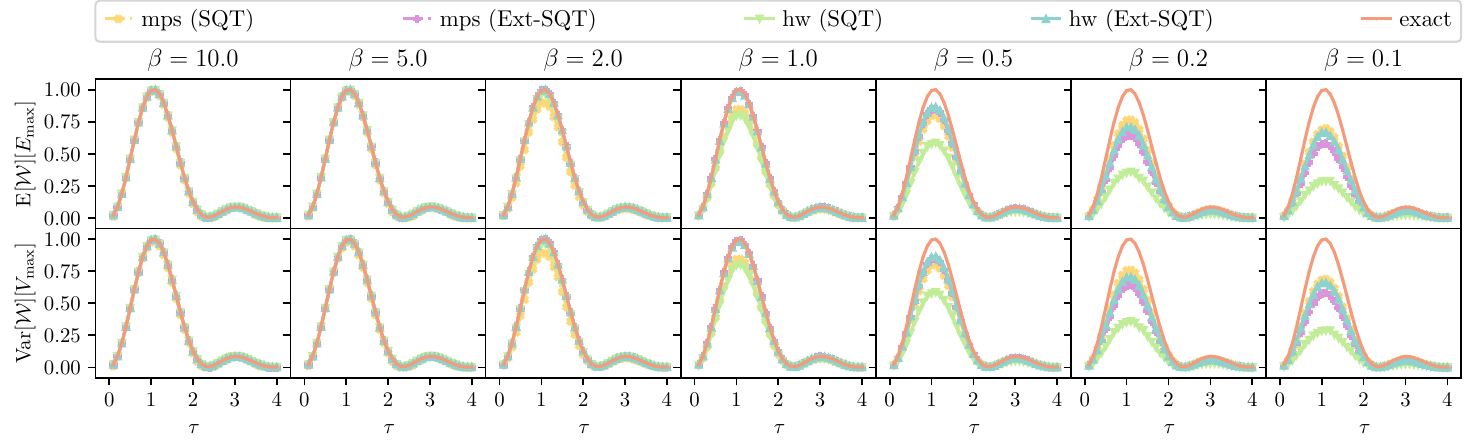}
\caption{\figuretitle{Error mitigation in the ``scan'' simulations.} First and second row: comparison between the mean $\meanwork$ and variance $\varwork$ of the quantum work distribution from quantum computing (yellow, purple circles) and tensor-network (green, blue triangles) data post-processed with the SQT and Ext-SQT techniques as a function of protocol duration for $\nqubits = 15$ qubits, $\ntrot = 10$ Trotter-Suzuki steps, and inverse temperatures $\beta = 10.0, 5.0, 2.0, 1.0, 0.5, 0.2, 0.1$ (left to right), taking results from exact calculations (red line) as reference.
}
\label{fig:mitigation_scan}
\end{figure}

In Fig.~\ref{fig:mitigation_scan}, we quantify the accuracy of post-processing for the results in Fig.~\ref{fig:figure_2_supplement} comparing $\meanwork$ and $\varwork$ from quantum hardware and tensor network samples against exact diagonalization at $\nqubits = 15$ qubits.
The ``SQT'' and ``Ext-SQT'' error mitigation techniques yield indistinguishable results at low temperature ($\beta > 2.0$, left). At intermediate temperature ($\beta=2.0, 1.0$) ``Ext-SQT'' outperforms ``SQT'' for both tensor network and quantum hardware data. The same is true at higher temperature for hardware data, which are affected by quantum noise, whereas for tensor network data ``SQT'' yields results in better agreement with ``Ext-SQT''.

\subsection{Simulations over values of $\gamma$ (``coupling'')}

\begin{figure}[h!]
\includegraphics[width=0.9\textwidth]{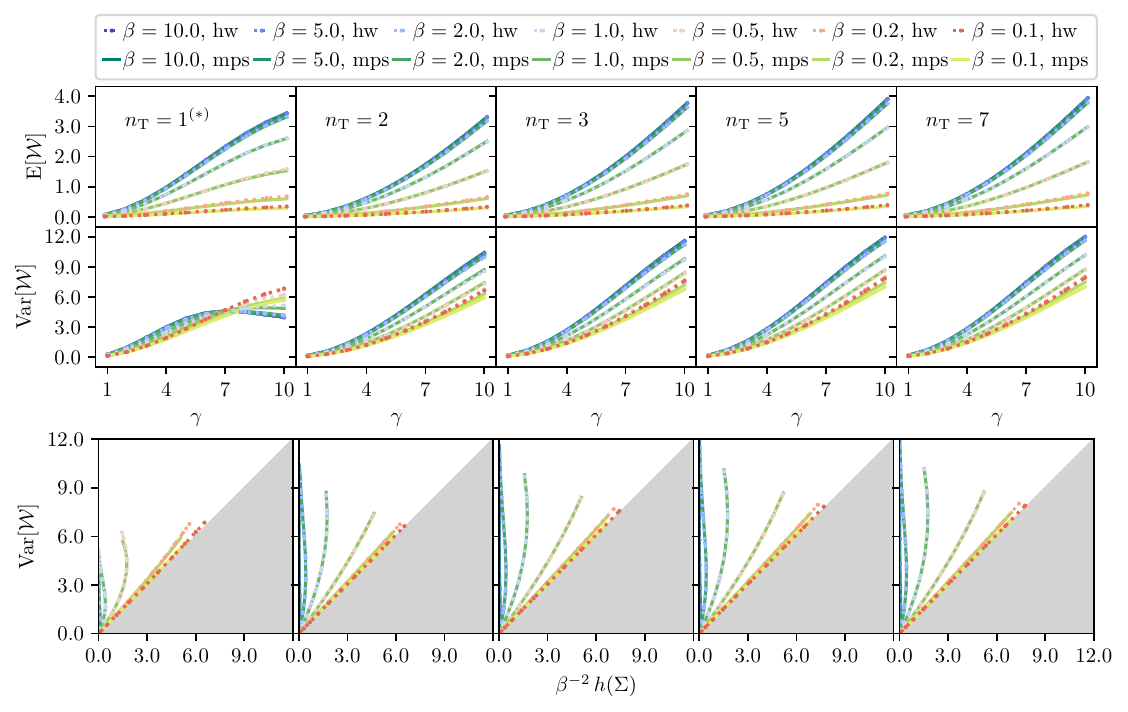}
\caption{\figuretitle{Trotter error in the ``coupling'' simulations.} First and second row: mean and variance of the quantum work distribution as a function of protocol strength $\gamma$, computed on \device{torino} (dotted curves, ``hw'') and with matrix product states (solid curves, ``mps'') for $\nqubits = 15$, $\ntrot = 1,3,5,7$ Trotter-Suzuki steps (left to right) and inverse temperatures from $\beta=10.0$ to $0.1$ (blue to red for ``hw'' and dark to light geen for ``mps''). Third row: TUR from the data in the first two rows, with violations occurring in the light gray region. Data with $\ntrot=1$ are divided by 2 to enhance visibility, and thus labeled with a (*) symbol.
}
\label{fig:trotter_coup}
\end{figure}

In Fig.~\ref{fig:trotter_coup}, we assess the impact of the Trotter approximation on the results in Fig.~\ref{fig:coup}. Results are essentially converged for $\ntrot=5$, with $\ntrot<3$ leading to significant quantitative and qualitative differences in the work distribution.

\begin{figure}[h!]
\includegraphics[width=0.9\textwidth]{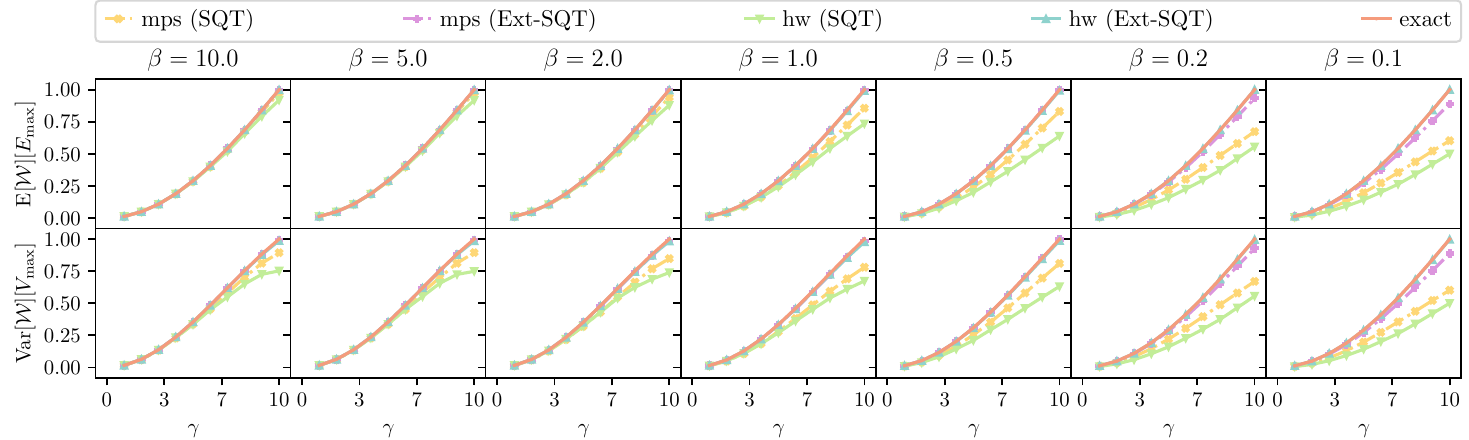}
\caption{\figuretitle{Post-processing in the ``coupling'' simulations.} First and second row: comparison between the mean $\meanwork$ and variance $\varwork$ of the quantum work distribution from quantum computing (yellow, purple circles) and tensor-network (green, blue triangles) data post-processed with the SQT and Ext-SQT techniques as a function of interaction strength, for $\nqubits = 15$ qubits, $\ntrot = 10$ Trotter-Suzuki steps, and inverse temperatures $\beta = 10.0, 5.0, 2.0, 1.0, 0.5, 0.2, 0.1$ (left to right), taking results from exact calculations (red line) as reference.
}
\label{fig:mitigation_coup}
\end{figure}

In Fig.~\ref{fig:mitigation_coup}, we quantify the effectiveness of post-processing on the results in Fig.~\ref{fig:coup} comparing $\meanwork$ and $\varwork$ from quantum hardware and tensor network samples against exact diagonalization at $\nqubits = 15$ qubits.
The trends seen in Fig.~\ref{fig:mitigation_scan} are confirmed here. As naturally expected, the differences between exact and post-processed results are more pronounced for higher values of $\gamma$, i.e., for stronger coupling between spins. The Ext-SQT technique is systematically more effective than the SQT technique, and the improvement due to the extension is more pronounced as $\gamma$ increases.


\subsection{Simulations scanning values of $\nqubits$}

\begin{figure}[h!]
\includegraphics[width=\textwidth]{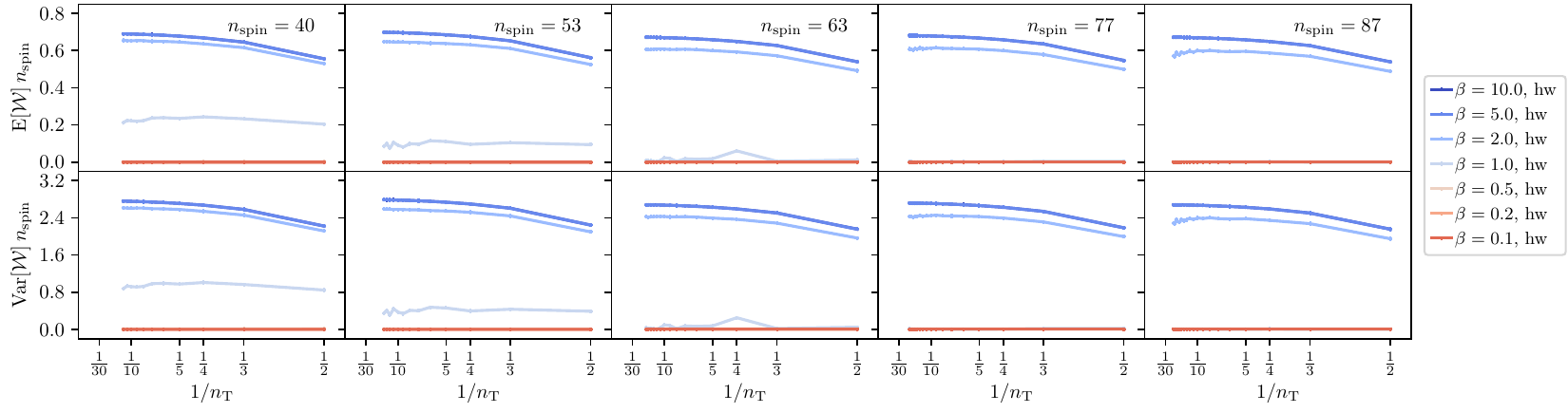}
\includegraphics[width=\textwidth]{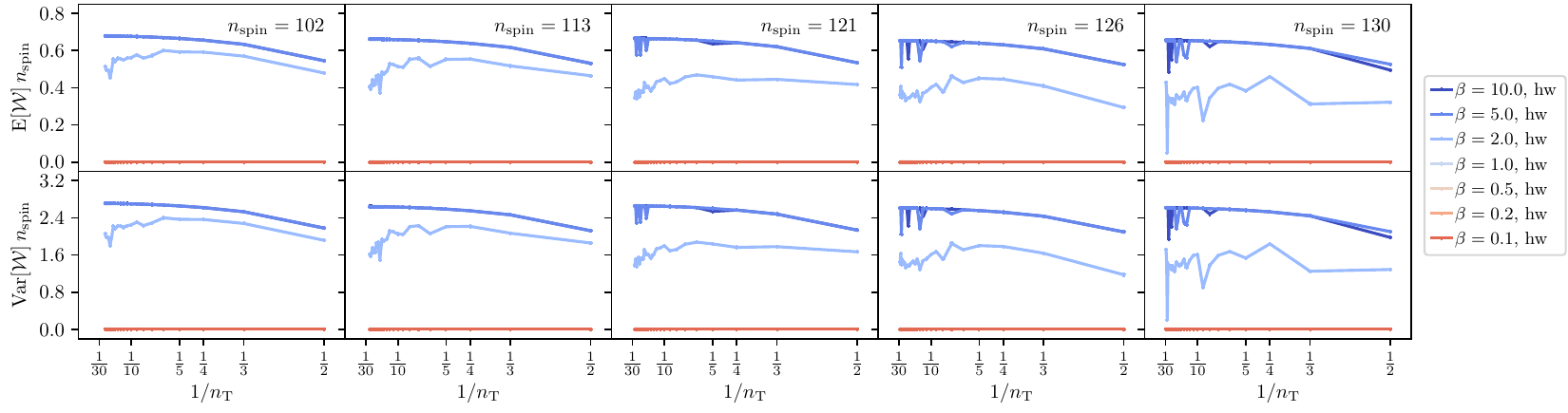}
\caption{\figuretitle{Trotter error in the ``size'' simulations.} First row: mean $\meanwork$ of the quantum work distribution as a function of the number of Trotter-Suzuki steps size computed on \device{torino} with Ext-SQT post-processing for $\nqubits = 40$ to $87$ qubits (left to right, top) and $\nqubits = 102$ to $130$ (left to right, bottom), time $\tau=1.0$, and inverse temperatures decreasing from $\beta = 10.0$ to $\beta=0.1$ (from blue to red).
Second row: variance $\varwork$ of the quantum work distribution under the same conditions as in the first row.}
\label{fig:trotter_size}
\end{figure}

In Fig.~\ref{fig:trotter_size}, we quantify the impact of the Trotter approximation on the results in Fig.~\ref{fig:size} of the main text. Results are essentially converged in number of Trotter-Suzuki steps at the highest value, corresponding to the diameter of the underlying Hamiltonian graph (leftmost point), with pronounced fluctuations occurring for more than $\ntrot \simeq 10$ Trotter-Suzuki steps for $\nqubits = 102$ to $130$.

\section{Extreme depolarizing channel analysis (``white noise'')}
\label{sec:appendix_white}

The raw data in Fig.~\ref{fig:size} show a mean and a variance of the work distribution that increase with system size, which is different from the prediction of response theory.
To understand this behavior, let us analyze the work distribution assuming that the drive protocol is corrupted by extreme depolarizing noise. This assumption consists in requiring that samples drawn at the beginning of the protocol (marked $\bts{x}$ in Fig.~\ref{fig:setup}) follow a Gibbs distribution, whereas those drawn at the end of the protocol (marked $\bts{y}$ in Fig.~\ref{fig:setup}) follow a uniform distribution,
\begin{equation}
\label{eq:white_noise_model}
p(\bts{y},\bts{x}) = 
\langle \bts{y} | \gibbs{0}{0} | \bts{y} \rangle 
\langle \bts{x} | \gibbs{0}{\beta} | \bts{x} \rangle
\;,
\end{equation}
where $\gibbs{0}{0} = \frac{I}{2^{\nqubits}}$ is the infinite-temperature state. The average work can be computed as
\begin{equation}
\begin{split}
&\meanwork 
= \sum_{\bts{x}\bts{y}} \left[ E_{\bts{y}} - E_{\bts{x}} \right] p(\bts{y},\bts{x}) 
= \sum_{xy} \Big[ 
\langle \bts{y} | H_0 | \bts{y} \rangle - 
\langle \bts{x} | H_0 | \bts{x} \rangle \Big] 
\langle \bts{y} | \gibbs{0}{0}   | \bts{y} \rangle 
\langle \bts{x} | \gibbs{0}{\beta} | \bts{x} \rangle \;.
\end{split}
\end{equation}
Recalling that for two diagonal operators $D_1$ and $D_2$ and for a generic configuration $\bts{z}$ one has $\langle \bts{z} | D_1 | \bts{z} \rangle \langle \bts{z} | D_2 | \bts{z} \rangle = \langle \bts{z} | D_1 D_2 | \bts{z} \rangle$, that $H_0$ and $\gibbs{0}{0}$ are diagonal operators, and that $H_0$ is also traceless, we obtain
\begin{equation}
\begin{split}
\meanwork &= 
\trace{ H_0 \gibbs{0}{0} } 
\trace{ \gibbs{0}{\beta} } 
- 
\trace{ \gibbs{0}{0} } \trace{ H_0 \gibbs{0}{\beta} } 
= - \sum_q \trace{ Z_q \frac{e^{\beta Z_q}}{\trace{e^{\beta Z_q}}} } = \nqubits \, \tanh(\beta) \;.
\end{split}
\end{equation}
The average work is therefore proportional to the system size.

A similar formula holds for the mean of the square of the work,
\begin{equation}
\begin{split}
&\mathrm{E}[\mathcal{W}^2] = 
\sum_{\bts{x}\bts{y}} \left[ E_{\bts{y}} - E_{\bts{x}} \right]^2 p(\bts{y},\bts{x}) 
=     \trace{ H_0^2 \gibbs{0}{0} } \trace{       \gibbs{0}{\beta} } 
+     \trace{       \gibbs{0}{0} } \trace{ H_0^2 \gibbs{0}{\beta} } 
-2 \, \trace{ H_0   \gibbs{0}{0} } \trace{ H_0   \gibbs{0}{\beta} } \\
\end{split}
\end{equation}
Recalling that Gibbs states have trace 1, that $H_0$ is traceless, that
\begin{equation}
\begin{split}
\trace{ H_0^2 \gibbs{0}{0} } 
&= \frac{1}{2^{\nqubits}} \sum_{pr} \trace{ Z_p Z_r } 
= \frac{1}{2^{\nqubits}} \sum_{pr} \delta_{pr} 2^{\nqubits} = \nqubits \;, \\
\trace{ H_0^2 \gibbs{0}{\beta} } 
&= \sum_{pr} \trace{ Z_p Z_r \frac{e^{\beta (Z_p+Z_r)}}{\trace{e^{\beta (Z_p+Z_r)}}} } 
= \sum_{pr} \delta_{pr} + \sum_{pr} (1-\delta_{pr}) \tanh(\beta)^2
= \nqubits + (\nqubits^2-\nqubits) \tanh(\beta)^2 \;,
\end{split}
\end{equation}
one readily obtains
\begin{equation}
\varwork = \mathrm{E}[\mathcal{W}^2] - \meanwork^2 = \nqubits \, \big[ 2 - \tanh(\beta)^2 \big]
\;.
\end{equation}
The variance of the work is also proportional to system size.

\begin{figure}[b!]
\includegraphics[width=0.95\textwidth]{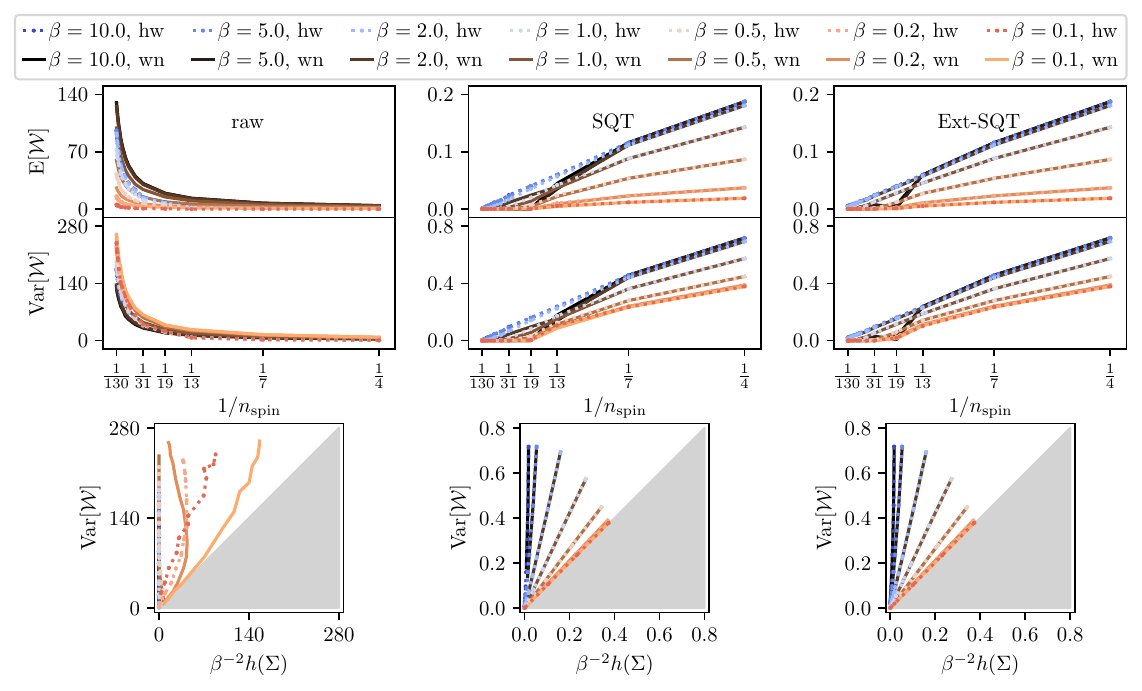}
\caption{\figuretitle{White noise analysis.}
First and second row: mean $\meanwork$ and variance $\varwork$ of the quantum work distribution as a function of system size computed on \device{torino} (dotted curves marked ``hw'') and with a combination of thermally distributed initial configurations ${\bf{x}}$ and uniformly distributed final configurations ${\bf{y}}$ (solid curves marked ``wn'') for $\nqubits = 4$ to $130$ qubits, time $\tau=1.0$, $\ntrot$ Trotter-Suzuki steps equal to the diameter of the underlying Hamiltonian graph, and inverse temperatures decreasing from $\beta = 10.0$ to $\beta=0.1$ (from blue to red for ``hw'' data and from black to copper for ``wn'' data, respectively). Results are raw (left) and post-processed with SQT and Ext-SQT techniques (center, right).
Third row: thermodynamic uncertainty relations from the data in the first and second row, with violations occurring in the gray region.
}
\label{fig:white_noise}
\end{figure}

These trends are reflected in Fig.~\ref{fig:white_noise}, where raw quantum computing data and data sampled from the distribution Eq.~\eqref{eq:white_noise_model} (labeled ``wn'' in the Figure) show $\meanwork$ and $\varwork$ increasing with system size. However, data following the distribution in Eq.~\eqref{eq:white_noise_model} overestimate the mean work compared to quantum computing data, for all values of the inverse temperature. Deviations are around $\simeq 35$ units at low temperature (blue dotted curve, black solid curve) and become more pronounced at higher inverse temperature, reaching $\simeq 50$ units at high temperature (red dotted curve, copper solid curve).
For $\varwork$, the situation is somewhat different: ``wn'' data underestimate the variance at low temperature and overestimate it at higher temperature.
These behaviors indicate that Eq.~\eqref{eq:white_noise_model} is not a quantitative model for the noise affecting the device, although it is useful to understand the scaling of $\meanwork$ and $\varwork$ with system size in raw quantum computing data.

Differences between data from quantum computing and ``wn'' data are also visible upon post-processing: for small values of $\nqubits$, results post-processed with the SQT method are in good agreement with each other. When $\nqubits$ increases, the agreement worsens, with the ``wn'' data underestimating $\meanwork$ and $\varwork$. As seen, the Ext-SQT method improves the ``wn'' data for $\nqubits = 13$ but leads to underestimating the momenta of the work distribution beyond that point. A similar phenomenon is also present in quantum computing data (see Fig.~\ref{fig:mitigation_scan}) and originates from the pruning operation documented in Fig.~\ref{fig:extended}, which is more aggressive in the presence of flatter distributions, e.g. at low inverse temperature and for ``wn'' data.

%\bibliography{main}


\end{document}